\documentclass{article}

\usepackage{arxiv}

\usepackage{makecell}
\usepackage[utf8]{inputenc} 
\usepackage[T1]{fontenc}    
\usepackage{hyperref}       
\usepackage{url}            
\usepackage{booktabs}       
\usepackage{amsfonts}       
\usepackage{nicefrac}       
\usepackage{microtype}      
\usepackage{graphicx}
\usepackage{doi}
\usepackage[numbers]{natbib}
\usepackage{setspace}
\usepackage{amsmath}
\usepackage{amssymb}
\usepackage{eucal}
\usepackage{tcolorbox}
\tcbset{
  myquestionbox/.style={
    colback=gray!5,
    colframe=black,
    boxrule=0.5pt,
    arc=2pt,
    left=4pt,
    right=4pt,
    top=4pt,
    bottom=4pt,
    fonttitle=\bfseries,
    before skip=8pt,
    after skip=8pt
  }
}
\usepackage[export]{adjustbox}

\newcommand{\step}[1]{%
  \tikz[baseline=(char.base)]{%
    \node[shape=circle,draw,inner sep=0.5pt] (char) {\footnotesize #1};}}

\usepackage{multirow}
\usepackage{booktabs}  
\usepackage{rotating}
\usepackage{soul,xcolor}

\usepackage{algorithm}
\usepackage{algpseudocode}
\usepackage{listings}
\usepackage{xspace}

\usepackage[table]{xcolor}
\usepackage{soul}
\usepackage{tabularx}

\newcolumntype{L}{>{\raggedright\arraybackslash}X}

\usepackage{algorithm}
\usepackage{algpseudocode}
\usepackage{listings}
\usepackage{xspace}

\def\tsc#1{\csdef{#1}{\textsc{\lowercase{#1}}\xspace}}
\tsc{WGM}
\tsc{QE}
\tsc{EP}
\tsc{PMS}
\tsc{BEC}
\tsc{DE}
\def\tsc#1{\csdef{#1}{\textsc{\lowercase{#1}}\xspace}}
\tsc{WGM}
\tsc{QE}
\tsc{EP}
\tsc{PMS}
\tsc{BEC}
\tsc{DE}

\begin{document}
\title{Physics-Grounded Multi-Agent Architecture for Traceable, Risk-Aware Human–AI Decision Support in Manufacturing}

\author{
Danny Hoang$^{1}$ \quad
Ryan Matthiessen$^{2}$ \quad
Christopher Miller$^{2}$ \quad
Nasir Mannan$^{2}$\\
\textbf{Ruby ElKharboutly}$^{3}$ \quad
\textbf{David Gorsich}$^{4}$ \quad
\textbf{Matthew P.~Castanier}$^{4}$ \quad
\textbf{Farhad Imani}$^{1}$\thanks{Corresponding author: farhad.imani@uconn.edu\\
DISTRIBUTION STATEMENT A. Approved for public release; distribution is unlimited. OPSEC9768}
\\[0.75em]
$^{1}$School of Mechanical, Aerospace, and Manufacturing Engineering, University of Connecticut, Storrs, CT, USA\\
$^{2}$Connecticut Center for Advanced Technology, East Hartford, CT, USA\\
$^{3}$Quinnipiac University, Hamden, CT, USA\\
$^{4}$DEVCOM Ground Vehicle Systems Center, Warren, MI, USA
}

\date{}

\renewcommand{\headeright}{}
\renewcommand{\undertitle}{}
\renewcommand{\shorttitle}{Physics-Grounded Agents for Trustworthy Manufacturing}

\hypersetup{
pdftitle={Physics-Grounded Multi-Agent Architecture for Traceable, Risk-Aware Human–AI Decision Support in Manufacturing},
pdfauthor={Danny Hoang, Ryan Matthiessen, Christopher Miller, Nasir Mannan, Ruby ElKharboutly, David Gorsich, Matthew P. Castanier, Farhad Imani},
pdfkeywords={Multi-Agent Systems, Industrial Large Language Models, Physics-Grounded AI, Traceable Decision Support
}
}
\maketitle

\begin{abstract}
High-precision CNC machining of free-form aerospace components requires bounded compensations informed by inspection, simulation, and process knowledge. Off-the-shelf large language model (LLM) assistants can generate text, but they do not reliably execute risk-constrained multi-step numerical workflows or provide auditable provenance for high-stakes decisions. We present multi-agent knowledge analysis (MAKA), a human-in-the-loop decision-support architecture that separates intent routing, tools-only quantitative analysis, knowledge graph retrieval, and critic-based verification that enforces physical plausibility, safety bounds, and provenance completeness before recommendations are surfaced for human approval. MAKA is instantiated on a Ti-6Al-4V rotor blade machining testbed by fusing virtual-machining path-tracking error fields, cutting-force and deflection simulations, and scan-based 3D inspection deviation maps from 16 blades. The analysis decomposes deviation into an evidence-linked pathing component, a drift-based wear proxy capturing systematic evolution across parts, a residual systematic compliance term, and a variability proxy for instability-aware escalation. In a three-level tool-orchestration benchmark (single-step through $\geq$3-step stateful sequences), MAKA improves successful tool execution by up to 87.5 percentage points relative to an unstructured single-model interaction pattern with identical tool access. Digital twin what-if studies show MAKA can coordinate traceable compensation candidates that reduce predicted surface deviation from order $10^{-2}$~in to approximately $\pm 10^{-3}$~in over most of the blade within the simulation environment, providing a pre-deployment verification signal for risk-aware human decision-making.
\end{abstract}

\section{Introduction}

Advanced computer numerical control (CNC) manufacturing systems are increasingly embedded in complex production environments in which cutting mechanics, structural compliance, servo control, fixturing, sensing, and part geometry interact in strongly nonlinear and state dependent ways~\cite{zhao2024unified,chen2024integrated,williams2025securing}. Achieving consistent part quality in this setting requires decisions that reconcile heterogeneous evidence streams under stringent tolerance, throughput, and safety constraints. Micrometer level requirements are often specified for critical free form surfaces, yet the limiting factor in many industrial workflows is not the availability of data. The limiting factor is the lack of integrated and accountable interpretation that converts multi source evidence into bounded actions with explicit justification. Conventional process planning pipelines and basic statistical monitoring typically treat operations or features in isolation and therefore provide no unified view of simulation outputs, inspection metrology, and prior machining knowledge~\cite{markatos2023manufacturing}. This fragmentation is increasingly misaligned with modern manufacturing practice where engineering decisions must be justified, auditable, and robust to uncertainty.

In standard industrial practice, tool paths are generated and verified offline in computer aided manufacturing (CAM) and virtual machining environments~\cite{liao2024data}, while process parameters are tuned using a combination of design of experiments studies, empirical rules, and shop floor experience~\cite{cao2025multisensor,mohanta2024experimental}. Quality control is then performed downstream via sparse inspection and statistical process control charts, which summarize deviations at a limited set of features but provide only coarse diagnostics of underlying error mechanisms~\cite{escobar2021quality}. These stages are typically connected through manual interpretation rather than through a formal evidence fusion and verification pipeline. As a result, rich multimodal information available in modern CNC manufacturing systems~\cite{hoang2025hyperdimensional,chen2025federated}, including deflection simulations, virtual machine path tracking predictions, dense scan derived deviation fields, machine telemetry, and documented best practices remains distributed across software tools and data silos. The absence of a consistent mechanism for coordinate consistent fusion, provenance tracking, and risk bounded recommendation makes it difficult to deploy automated decision support that humans can rely on in high consequence settings.

These limitations become more pronounced for non prismatic, free-form components where local geometry~\cite{wang2023toolpath}, tool engagement~\cite{singh2025curvature}, and fixture stiffness vary continuously along the tool path~\cite{li2023part}. In such settings, offline one shot planning struggles to keep up with variability introduced by changes in machines, tooling, materials, and fixturing configurations across production systems. Even small modifications to any of these elements can invalidate previously calibrated parameter sets and degrade transferability of best settings between nominally similar setups. Engineers are therefore often forced into iterative trial and error adjustment of offsets, tool paths, and parameters~\cite{kukreja2023optimal,garba2025enhancing}. At the same time, attributing observed geometric deviation to specific contributors such as path tracking error, tool wear, or workpiece compliance requires combining imperfect models and noisy measurements~\cite{chen2024study}. Incorrect attribution directly leads to incorrect compensation and costs that include scrap, rework, tool damage, and schedule disruption.

Large language models (LLMs) offer capabilities that are attractive for industrial decision support, including natural language interaction for operators~\cite{yao2025evaluation,hoang2025knowledge}, automated organization and summarization of heterogeneous artifacts~\cite{chen2025multi,chen2025can}, and coordination across multiple software tools through tool calling patterns~\cite{qu2025tool,xu2025llm}. In principle, an LLM enabled assistant could read simulation reports~\cite{li2025mechrag}, interpret metrology outputs~\cite{fan2025mavila}, consult machining studies, and propose targeted updates to tool paths and process parameters~\cite{jeon2025chatcnc}. In practice, off the shelf LLMs are not designed for high stakes industrial workflows. They lack explicit representations of machine kinematics, cutting mechanics, and CAM and CNC constraints, are vulnerable to hallucination and inconsistent numerical reasoning, and provide weak guarantees of provenance, repeatability, and bounded action, which are central requirements for trustworthy human AI collaboration in production environments~\cite{chen2025large}. These limitations are amplified by resource and security constraints at or near the edge, where extensive retraining and unconstrained multimodal reasoning are impractical. Effective human AI symbiosis in CNC manufacturing therefore requires architectures that constrain machining critical computation to deterministic tools, retrieve domain knowledge with traceable evidence, and verify candidate actions against physical plausibility and safety bounds before they are presented for human decision making.

We introduce the multi agent knowledge analysis (MAKA) framework as an LLM driven agentic decision support layer for CNC manufacturing systems. Rather than treating the LLM as a monolithic generator of machining strategies, MAKA decomposes the evidence to recommendation pipeline into specialized auditable roles aligned with risk aware industrial symbiosis. A Central agent performs intent interpretation and task routing. An Analysis agent is constrained to tools only quantitative computation. A Knowledge Graph agent performs evidence linked retrieval over domain knowledge constructs. A Critic agent evaluates whether candidate outputs are coherent, physically plausible, within safety bounds, and supported by explicit provenance before they are surfaced for human review. This separation of concerns supports accountable operation because machining critical quantities are produced through deterministic tool calls and traceable retrieval rather than through unconstrained text generation. By offloading bound checking, consistency checks, and provenance completeness to the Critic agent, MAKA reduces the cognitive burden of multi modal evidence synthesis and enables the human expert to focus on higher level tradeoffs such as quality, tool life, and cycle time.

We instantiate MAKA on an aerospace application, the production of Ti-6Al-4V titanium rotor blades on industrial CNC equipment. MAKA fuses evidence across the manufacturing digital thread, including virtual machining outputs for path tracking deviation, cutting force and workpiece deflection simulations, and dense scan based inspection deviation maps acquired across multiple manufactured blades, together with a machining knowledge graph derived from a curated corpus of technical literature capturing machining mechanisms and constraints across multiple CNC manufacturing systems. The agents structure this evidence into interpretable deviation attributions including pathing, a tool wear proxy, and a residual compliance term, then synthesize candidate compensation strategies expressed as tool path modifications, tool length and radius offsets, and process parameter adjustments with provenance linked explanations and verification checks. The contributions of this work are threefold. First, we present a physics grounded adaptation of LLM based decision support for CNC manufacturing via explicit tool interfaces and knowledge graph integration that operationalizes accountable human oversight. Second, we propose an agent based architecture that improves reliability of multi step industrial tool orchestration under resource constraints through separation of intent routing, computation, retrieval, and verification. Third, we demonstrate within an industrial digital twin toolchain, using virtual machining and physics based simulation environments, that multimodal data fusion combining simulation outputs, scan derived deviation fields, and symbolic machining knowledge can be executed within a single traceable decision support workflow that provides a verified pre deployment baseline for risk aware human decision making.

The remainder of the paper is organized as follows. Section~\ref{sec:background} brief review of related work in manufacturing knowledge systems, LLM-based decision support, CNC automation, and geometric error modeling, along with their limitations. Section~\ref{sec:methodology} presents the MAKA architecture as an agentic layer for CNC manufacturing systems, detailing the roles of the central, analysis, knowledge, and critic agents and the associated agentic tools. Section~\ref{sec:experimental_design} describes the experimental setup used to evaluate MAKA on rotor blade machining. Section~\ref{sec:results} reports the results on physics-based deviation decomposition, automated parameter selection, and toolpath compensation. Finally, Section~\ref{sec:conclusion} discusses limitations and opportunities for deploying such agentic LLM frameworks in broader CNC manufacturing environments.


\section{Research Background}
\label{sec:background}

In industrial systems, human--AI decision support is not achieved by adding a conversational interface alone; it requires that the AI system (i) integrate heterogeneous evidence across the manufacturing digital thread, (ii) remain grounded in validated physics and deterministic computation where required, (iii) expose provenance and uncertainty in forms that support human oversight, and (iv) enforce bounded autonomy through verification and safety checks before recommendations are acted upon. We organize the background along four main axes: (i) knowledge graph, ontology, and skill-based infrastructures for representing manufacturing knowledge capabilities; (ii) LLM-based knowledge assistants and multi-agent frameworks for tool-using decision support in production systems; (iii) LLM-based CNC programming and monitoring, where reliability and verification are central; and (iv) data-driven models of geometric error in aero-engine blades and related thin-walled, free-form parts. 

A long-standing response to fragmentation in manufacturing knowledge is to formalize process knowledge and system capabilities into explicit, machine-interpretable structures. Knowledge graph-based process planning frameworks encode parts, operations, resources, and precedence constraints as symbolic graphs and use graph reasoning to support computer-aided process planning, reuse, and consistency checking~\cite{xiao2023knowledge}. These representations help capture best practices and make dependencies explicit, which is valuable for interoperability and lifecycle management; however, they typically operate at the level of discrete operations and resources rather than spatially resolved process physics, and they rarely represent dense simulation fields, scan-derived deviation maps, or uncertainty in a way that directly supports high-stakes compensation decisions on free-form geometries. In parallel, skill-based architectures for production systems expose machine capabilities as standardized skills that can be orchestrated at runtime across heterogeneous equipment~\cite{hossfeld2024universal}. Skill-based approaches advance modularity and reconfigurability, but they often treat the internal machining process as a black box and provide limited support for fusing simulation, metrology, and expert heuristics into an auditable causal account of why a part deviates and which bounded interventions are appropriate. This creates a gap between what a system can do (skills, operations) and why a particular action is recommended (evidence, provenance, physical plausibility), a gap that becomes acute when decisions must be justified to human experts and recorded for accountability.

More recently, LLMs have been explored as a way to unlock unstructured manufacturing knowledge and provide conversational interfaces to complex production systems. Survey work on LLMs for manufacturing highlights opportunities across the product lifecycle, including design assistance, documentation mining, and decision support, and emphasizes retrieval augmented generation (RAG) and tool use as key patterns for integrating domain knowledge and software systems~\cite{li2024large}. Empirical studies of LLM-powered knowledge sharing on the shop floor demonstrate that operators can query manuals, procedures, and historical reports more effectively than with keyword search alone~\cite{kernan2024knowledge}. These contributions are important for human-centric industrial interaction, but most remain predominantly text-centric; they reason over documents and structured tables rather than over the multimodal artifacts that dominate machining decisions (e.g., CAM outputs, virtual machining fields, scan-derived deviation maps, or CNC telemetry). Moreover, they typically do not enforce bounded autonomy or provide systematic mechanisms for provenance completeness (linking each numeric or procedural recommendation to a verifiable source) and risk-aware verification (rejecting or escalating recommendations that violate physical plausibility or safety constraints).

A complementary strand of work embeds LLMs into multi-agent or agent-based manufacturing architectures. In such systems, language models interpret human instructions, decompose tasks, and coordinate agents that ultimately execute operations and may emit machine control programs~\cite{lim2024large}. Other frameworks translate informal textual specifications into structured process plans using LLM-based pipelines for task decomposition and workflow synthesis, sometimes combined with verification layers (e.g., state-machine checking) to ensure basic logical correctness before deployment on industrial platforms~\cite{ni2025large}. These approaches are promising for interoperability and adaptive autonomy, but their support for risk-aware symbiosis is often limited by two factors: (i) numerical and physical computations are still frequently performed within the language model loop without strong deterministic guarantees, and (ii) verification focuses on syntactic or logical consistency rather than on physically grounded plausibility, bounded action constraints, and provenance completeness over heterogeneous manufacturing evidence.

To improve numerical reliability and transparency, several systems couple LLMs with structured manufacturing knowledge. Knowledge graph-augmented frameworks build machining knowledge graphs from technical documents, vendor datasheets, and prior process plans, and use RAG over these graphs to answer CNC process planning queries with traceable numerical recommendations~\cite{hoang2025knowledge}. Multi-agent retrieval frameworks further decompose retrieval, ranking, and answer generation into specialized agents to integrate information from diverse manufacturing sources~\cite{mao2025multi}. In mechanical engineering more broadly, multimodal systems such as MechRAG integrate CAD and computer aided engineering (CAE) assets into LLM responses, enabling conversational access to simulation results and model artifacts~\cite{li2025mechrag}. CNC-focused assistants like ChatCNC attach LLMs to real-time machine data streams to provide conversational monitoring and diagnostics~\cite{jeon2025chatcnc}. Collectively, these works demonstrate that LLMs can be grounded in structured knowledge, multimodal engineering artifacts, and live telemetry. Nonetheless, they often stop short of (i) performing physics-aware decomposition of geometric error into interpretable contributors under explicit assumptions, (ii) orchestrating deterministic analysis tools through an auditable tool-call interface, and (iii) enforcing a verification gate that manages risk by bounding recommendations and escalating ambiguous cases to human experts.

\begin{table*}[t]
\centering
\small
\caption{Capabilities of representative related work compared with MAKA framework.}
\label{tab:related_work_maka}
\resizebox{\textwidth}{!}{
\begin{tabular}{cccccccc}
\toprule
\textbf{Model} &
\textbf{KG / Skills} &
\textbf{LLM} &
\textbf{Multi Agent} &
\textbf{Physics Decomposition} &
\textbf{Digital Twin Verification} &
\textbf{Tool Orchestration} \\
\midrule
Xiao et al. ~\cite{xiao2023knowledge} &
$\checkmark$ & $\times$ & $\times$ & $\times$ & $\times$ & $\times$ \\

Hossfeld \& Wortmann ~\cite{hossfeld2024universal} &
$\checkmark$ & $\times$ & $\times$ & $\times$  & $\times$ & $\times$ \\

Li et al.\ ~\cite{li2024large} &
$\times$ & $\checkmark$ & $\times$ & $\times$ & $\times$ & $\times$ \\

Freire et al.\ ~\cite{kernan2024knowledge} &
$\times$ & $\checkmark$ & $\times$ & $\times$ & $\times$ & $\times$ \\

Lim et al.\ ~\cite{lim2024large} &
$\times$ & $\checkmark$ & $\checkmark$ & $\times$ & $\times$ & $\checkmark$ \\

Ni et al.\ \cite{ni2025large} &
$\times$ & $\checkmark$ & $\times$ & $\times$ & $\times$  & $\checkmark$ \\

Hoang et al.\ ~\cite{hoang2025knowledge} &
$\checkmark$ & $\checkmark$ & $\times$ & $\times$ & $\times$ & $\times$ \\

Mao et al.\ ~\cite{mao2025multi} &
$\times$ & $\checkmark$ & $\checkmark$ & $\times$  & $\times$ & $\checkmark$ \\

Li \& Corney ~\cite{li2025mechrag} &
$\times$ & $\checkmark$ & $\times$ & $\times$ & $\times$ & $\times$ \\

Jeon et al.\ ~\cite{jeon2025chatcnc} &
$\times$ & $\checkmark$ & $\times$ & $\times$ & $\times$ & $\times$ \\

Šket et al.; Abdelaal et al.\ ~\cite{vsket2025large,abdelaal2025gllm} &
$\times$ & $\checkmark$ & $\times$ & $\times$ & $\times$ & $\times$ \\

Yang et al.\ ~\cite{yang2025aero} &
$\times$ & $\times$ & $\times$ & $\checkmark$ & $\times$ & $\times$ \\

\textbf{MAKA (Ours)} &
$\checkmark$ & $\checkmark$ & $\checkmark$ & $\checkmark$ & $\checkmark$ & $\checkmark$ \\
\bottomrule
\end{tabular}%
}
\end{table*}

At the same time, several studies have evaluated or proposed LLM-based generation of CNC programs directly from natural language descriptions or simple geometric sketches, highlighting both the promise and the risks of treating toolpath synthesis as a pure text generation task~\cite{jignasu2023towards,jeon2025chatcnc}. Comparative evaluations show that general purpose LLMs can emit syntactically valid G-code, but that the resulting programs often require nontrivial human correction and verification, particularly for complex geometries or nonstandard machine configurations~\cite{vsket2025large}. More specialized frameworks introduce domain-specific fine-tuning, retrieval, and self-correction loops, along with syntactic and geometric validation~\cite{abdelaal2025gllm}, yet they still operate largely outside of physics-based digital twin environments and often rely on a single language model to make intertwined geometric, physical, and process decisions. The AI system should not be treated as an unconstrained generator of machining strategies, but as an orchestrator of validated planning and physics tools, equipped with explicit verification and provenance mechanisms that reduce automation bias and support accountable human approval.

In parallel to these LLM-centric developments, there is a rich body of work on data-driven modeling of error distributions in thin-walled aero-engine blades and similar components. For example, Gaussian-mixture-model-based frameworks have been proposed to predict full-field milling error distributions for blades and impellers across different process conditions, using experimental data to characterize complex spatial patterns of deviation~\cite{yang2025aero}. Such models capture important aspects of the underlying process behavior and can provide informative priors for compensation. However, they are usually developed as standalone prediction tools, with limited integration into broader decision-making pipelines that also incorporate virtual machining, physics-based deflection simulation, and symbolic knowledge of machining constraints. They also rarely expose an evidence-linked explanation of how contributors such as path-tracking error, tool wear (or wear proxies), and compliance effects combine to yield observed deviation patterns on a specific production system. 

Table~\ref{tab:related_work_maka} summarizes representative contributions, emphasizing their implementation focus and the resulting limitations for risk-aware decision support on complex free-form geometries. As the table indicates, prior approaches either (i) emphasize symbolic knowledge representations modularity without deeply engaging with multi-modal machining physics and inspection evidence, (ii) focus on text-centric or monitoring-oriented use of LLMs that do not translate detailed geometric analysis into bounded, auditable modifications to tool paths or offsets, or (iii) treat CNC programming itself as a text generation problem with limited physics grounding and incomplete reliability guarantees. 

\section{Methodology}\label{sec:methodology}

\begin{figure*}
    \centering
    \includegraphics[width=\linewidth]{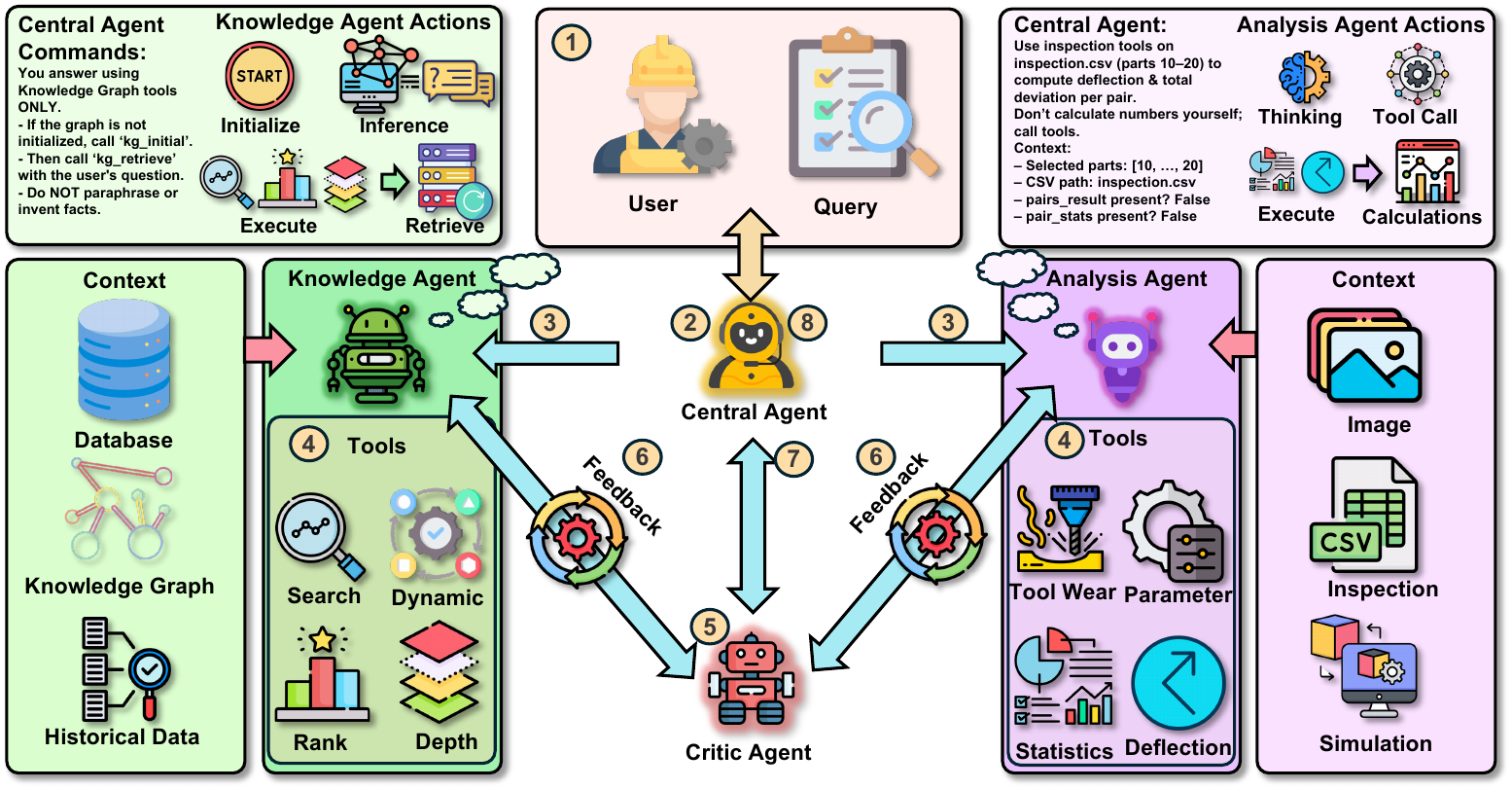}
    \caption{Multi-agent knowledge framework for CNC manufacturing. The Central, Knowledge Graph, Analysis, and Critic agents collaborate in an eight-step loop: (1) user query, (2 to 3) central planning and routing, (4) downstream analysis and knowledge retrieval, (5 to 7) critic evaluation and refinement, and (8) recommendation returned for human review and downstream CAM implementation.}
    \label{fig:methodology}
\end{figure*}

The multi agent knowledge analysis framework comprises four agents (Central, Knowledge Graph, Analysis, and Critic) that collaborate to answer and execute decision-support queries arising within a CNC manufacturing workflow. Figure~\ref{fig:methodology} summarizes the procedure as an eight-step loop. \step{1} A natural language query is provided by a user or a higher-level planning system. \step{2} The Central agent (CA) interprets the query in the context of the current manufacturing state, such as the active machine, toolpath, datasets, and previously computed summaries. \step{3} The CA constructs a plan and routes sub-tasks to either the Analysis agent (AA) for tool-grounded quantitative computation or the Knowledge Graph agent (KGA) for evidence retrieval from the machining knowledge graph. \step{4} The downstream agents execute their sub-tasks using deterministic tools and retrieval operators to produce intermediate results, such as deviation attributions, compensation candidates, and evidence-linked constraints. \step{5} These candidate outputs are passed to the Critic agent (CRA). \step{6} The CRA checks logical coherence, physical plausibility, compliance with safety bounds, and provenance completeness. \step{7} If any check fails, the CRA issues a revised instruction for refinement or escalates to the human when verification is insufficient. \step{8} Otherwise, the accepted recommendation and its audit trail are returned to the CA and surfaced to the user for review prior to downstream CAM implementation.
This loop implements an iterative decision-support workflow from a natural language intent to tool-grounded analysis and traceable recommendations. The subsections below define each agent and the constraints that enforce deterministic computation, evidence grounding, and risk-aware escalation.

\subsection{Central Agent}

The Central agent is the interface between the user and downstream agents. At turn $t$, the CA receives the user query $q_t$ and the internal state $s_t$, where $s_t$ contains references to loaded resources, cached intermediate results, and agent invocation history. Let $\mathcal{Q}$ denote the space of user queries, $\mathcal{S}$ the state space, $\mathcal{A}$ the set of available downstream agents, and $\tilde{\mathcal{Q}}$ the space of structured agent-specific instructions. The CA implements a routing policy
\begin{equation}
  (a_t,\tilde q_t,s_t^{+})=\pi_{C}(q_t,s_t)
\end{equation}
where $a_t\in\mathcal{A}$ specifies the next agent, $\tilde q_t\in\tilde{\mathcal{Q}}$ is a structured instruction tailored to that agent, and $s_t^{+}\in\mathcal{S}$ is the updated state.
In implementation, $\pi_{C}$ combines a language-model router with schema validation and a deterministic fallback. The router is constrained to emit a JSON object with fields that include the selected agent identifier, a normalized instruction string, required input references (files, part indices, tool identifiers), and an explicit list of requested tool categories. A JSON-schema validator defines the condition under which the output is considered valid. If validation fails, the CA falls back to a deterministic heuristic $\pi_{D}$ based on task keywords and available resources. Prior to routing, the CA applies a preprocessing update $s_t^{p}=u_{C}(q_t,s_t)$ to extract structured entities (e.g., part ranges, file paths, reset flags) so routing is evaluated on $(q_t,s_t^{p})$.

\subsection{Analysis Agent}

The Analysis agent performs data- and tool-centric reasoning. At turn $t$, the AA receives the routed instruction $\tilde q_t$, the current state $s_t$, and external resources $r_t$ (data handles, tool APIs, simulation artifacts). The AA is constrained so that any non-trivial numerical quantity included in its output must be produced by deterministic tools. Let $\mathcal{T}=\{T_1,\dots,T_M\}$ be the tool library. The AA selects a finite tool-call sequence $\tau_{1:K}=((j_1,x_1),\dots,(j_K,x_K))$ and executes it to obtain outputs $z_k=T_{j_k}(x_k)$, $k=1,\dots,K$. The AA returns a structured result $y_t^{A}$ together with an updated state $s_{t+1}$ that caches reusable artifacts and provenance metadata.

To enforce tool grounding, $y_t^{A}$ includes (i) a list of tool calls with arguments, (ii) tool outputs, and (iii) a provenance map that assigns each reported scalar or vector quantity to a specific tool output identifier. The CRA (Section~\ref{sec:critic_updated}) verifies that any numeric quantity in the final recommendation can be traced to this provenance map.

\subsection{Knowledge Graph Agent}

The Knowledge Graph agent answers conceptual, relational, and constraint queries over a symbolic graph $\mathcal{G}_t=(\mathcal{V}_t,\mathcal{E}_t,\mathcal{R}_t)$ of entities, typed relations, and labeled edges. Given $\tilde q_t$ and $s_t$, the KGA performs retrieval to build a context $c_t=\mathcal{R}_{K}(\tilde q_t,\mathcal{G}_t)$ and then produces an answer $y_t^{K}$ using the retrieved evidence together with any relevant observations carried in $s_t$.

Retrieval uses semantic embedding matching over graph triples and associated textual contexts. Let $f(\cdot)\in\mathbb{R}^d$ be a shared embedding model. Each stored item $i$ is a quadruple $\tau_i=(s_i,r_i,o_i,c_i)$ with a triple text representation and an associated context $c_i$. For query embedding $\mathbf{q}_t=f(\tilde q_t)$, each item receives a combined score
$s_i=\cos(\mathbf{q}_t,\mathbf{v}_i)+\lambda\,\cos(\mathbf{q}_t,\mathbf{u}_i)$ where $\mathbf{v}_i$ embeds the concatenation $(s_i,r_i,o_i,c_i)$, $\mathbf{u}_i$ embeds $c_i$, and $\lambda\in[0,1]$ weights contextual alignment. To control computation, an initial pre-pool $\mathcal{P}$ is formed by selecting the top $p=\min(N,\max(m,\lfloor \alpha N\rfloor))$ items under the base triple similarity $\cos(\mathbf{q}_t,\mathbf{v}_i)$, where $N$ is the number of stored items, $\alpha$ is a pruning fraction, and $m$ is a minimum retained count. Within $\mathcal{P}$, a data-adaptive inclusion floor $\tau=\mu+z\sigma$ is computed from the mean $\mu$ and standard deviation $\sigma$ of $\{s_i:i\in\mathcal{P}\}$. The initial candidate set $\mathcal{C}_0$ is formed from items with $s_i\ge\tau$, with a fallback to a fixed top-$k_0$ if empty and with $k$ bounded within $[k_{\ell},k_u]$. To capture multi-hop evidence, a bounded neighborhood expansion is applied to $\mathcal{C}_0$ using depth $d_M$ and a capped beam width $b_M$, with a visited set to prevent revisiting the same triple. The retrieved context $c_t$ is the set of selected triples and their supporting contexts.

For synthesis, the KGA returns (i) a structured answer and (ii) an evidence list of triple identifiers that support each claim. The CRA verifies that claims presented as constraints or best practices are supported by retrieved evidence identifiers, otherwise they are flagged for refinement or escalation.

\subsection{Critic Agent}
\label{sec:critic_updated}

The Critic agent evaluates candidate outputs from downstream agents and enforces risk-aware acceptance criteria before recommendations are surfaced. After an agent produces a candidate answer $y_t$, the CRA receives $(q_t,y_t,s_t)$ and returns a decision together with a possible refinement instruction. We model the critic output as
\begin{equation}
    (d_t,a_{t+1},\tilde q_{t+1},s_t^{+})=\pi_{R}(q_t,y_t,s_t),
\end{equation}
where $d_t\in\{\mathrm{accept},\mathrm{revise},\mathrm{escalate}\}$. If $d_t=\mathrm{accept}$, the system returns $y_t$ together with its audit trail. If $d_t=\mathrm{revise}$, the CRA selects $a_{t+1}\in\mathcal{A}$ and emits $\tilde q_{t+1}$ for refinement. If $d_t=\mathrm{escalate}$, the system returns a structured report that includes the candidate recommendation, the failed checks, and the minimal additional information required from a human operator or engineer.

The CRA evaluates four classes of checks: intent alignment with $q_t$, tool-grounding completeness (every numeric quantity must map to an AA tool output identifier), evidence grounding for retrieved constraints (every KG claim must cite retrieved triple identifiers), and safety checks that enforce physical plausibility and bounded action constraints (for example, maximum allowable offset magnitudes and sign consistency with declared coordinate conventions). A scalar quality score $J(q_t,y_t,s_t)\in[0,1]$ may be computed internally, but acceptance is not determined by $J$ alone; acceptance requires that the explicit checks above pass.

To guarantee termination, a critic budget $L\in\mathbb{N}$ limits the number of refinement iterations per user query. Let $n_t$ denote the number of CRA invocations for the current query. If $n_t\ge L$ and the candidate answer has not satisfied the acceptance criteria, the CRA returns $d_t=\mathrm{escalate}$ rather than forcing acceptance. In deployments that support interactive symbiosis, the CRA decision can be overridden by an explicit human approval signal, which is recorded as part of the audit trail.

\subsection{Agentic Tools}
\label{subsec:agent_tools}

Within the rotor blade case study, the MAKA system predicts and compensates geometric deviation by integrating three classes of evidence: virtual machining path-tracking error, inspection-derived deviation fields, and process knowledge constraints. The analysis separates systematic contributors that can be linked to explicit evidence from a residual term that captures compliance and other unmodeled systematic effects. The tool library  as shown in Table \ref{tab:agent_tools} exposes deterministic functions that load data, compute statistics, map simulation fields to inspection locations, estimate drift and variability proxies from multi-part data, and compute compensation offsets under a simplified tool-geometry model. While the tool implementation in this case study utilizes rotor blade specific nomenclature (e.g., rb prefixes), the MAKA architecture treats these as modular skills adhering to a generic geometric interface. The agents reason over standardized inputs (point clouds, deviation vectors, metadata) rather than hard-coded geometry logic. Consequently, adapting this framework to a different component (e.g., an engine casing) requires only the substitution of the underlying computational libraries (the tool layer) without retraining the agentic reasoning or restructuring the orchestration logic.

\noindent\textbf{Inspection and simulation alignment.}
For each inspection pair $k$ and part $n\in\{1,\dots,N\}$, let $\delta^{\mathrm{P}}_{k,n}$ and $\delta^{\mathrm{S}}_{k,n}$ denote the signed normal deviations measured on the pressure and suction surfaces after scan-based registration to the nominal CAD model. The paired measurement is summarized as the combined thickness-direction deviation $v_{k,n}=\delta^{\mathrm{P}}_{k,n}+\delta^{\mathrm{S}}_{k,n}$. Under the paired-surface sign convention, the corresponding per-surface deviation is $s_{k,n}=v_{k,n}/2$. The tool \texttt{compute\_inspection\_pairs} constructs the set of matched pairs and provides $\{\delta^{\mathrm{P}}_{k,n},\delta^{\mathrm{S}}_{k,n}\}_{k,n}$ together with the derived $v_{k,n}$ and $s_{k,n}$. The tool \texttt{fetch\_inspection\_slices} returns selected subsets and caches normalized representations.

Run MyVirtual Machine provides a simulated path-tracking error field that is projected onto the same inspection keys $k$ through a deterministic mapping operator that aligns simulation outputs and inspection locations in the CAD frame. Let $r_k$ denote the simulated combined deviation at pair $k$ as $v_{k,n}$, and define the per-surface pathing deviation as $p_k=r_k/2$. The tool \texttt{rb\_compute\_pathing\_dev} implements this projection and returns $p_k$.

\noindent\textbf{Multi-part decomposition using drift and variability proxies.}
For each pair $k$, define the non-pathing deviation for part $n$ as $u_{k,n}=s_{k,n}-p_k$. To separate systematic drift from dispersion, we fit a per-pair affine model over part index,
$u_{k,n}=c_k+b_k(n-1)+\epsilon_{k,n}$,
where $c_k$ is a baseline systematic term, $b_k$ is a drift rate per part, and $\epsilon_{k,n}$ is a zero-mean residual. The least-squares drift estimator is
$b_k=\frac{\sum_{n=1}^{N}(n-\bar n)(u_{k,n}-\bar u_k)}{\sum_{n=1}^{N}(n-\bar n)^2}$,
with $\bar n=(N+1)/2$ and $\bar u_k=\frac{1}{N}\sum_{n=1}^{N}u_{k,n}$, and the baseline term is $c_k=\bar u_k-b_k(\bar n-1)$. The drift magnitude at the last observed part is $w_k^{d}=b_k(N-1)$. The residual dispersion is quantified as
$w_k^{v}=\sqrt{\frac{1}{N-2}\sum_{n=1}^{N}\epsilon_{k,n}^2}$ for $N\ge 3$, where $\epsilon_{k,n}=u_{k,n}-c_k-b_k(n-1)$.

In this decomposition, $w_k^{d}$ is a drift proxy that is consistent with systematic evolution across part index, and $w_k^{v}$ is a variability proxy that captures process instability and unmodeled fluctuations. The methodology does not equate $w_k^{v}$ with a physical wear state. Instead, $w_k^{v}$ is used as a risk indicator for verification and escalation. The tools \texttt{rb\_compute\_wear\_drift} and \texttt{rb\_compute\_process\_variability} implement these estimators, while \texttt{rb\_compute\_residual\_systematic} returns the baseline term $c_k$.

For reporting and downstream reasoning, the mean surface deviation is $\bar s_k=\frac{1}{N}\sum_{n=1}^{N}s_{k,n}$ and the predicted surface deviation for a target part index $n^\star$ is
$\hat s_k(n^\star)=p_k+c_k+b_k(n^\star-1)$.
The residual term $c_k$ is interpreted as a residual systematic compliance term that includes stiffness-related effects and other systematic deviations not captured by rigid-body path-tracking simulation and drift. It is not uniquely identifiable as deflection alone.

\noindent\textbf{Stable attribution ratios for decision support.}
To avoid numerical instability in ratio metrics, attribution uses magnitudes and an $\epsilon$-floor. For a target part $n^\star$, define the component magnitudes
$p_k^{a}=|p_k|$, $c_k^{a}=|c_k|$, and $w_k^{a}=|b_k(n^\star-1)|$, and define the total magnitude
$a_k=p_k^{a}+c_k^{a}+w_k^{a}+\epsilon$ with small $\epsilon>0$. The corresponding attribution fractions are
$\phi_k^{p}=p_k^{a}/a_k$, $\phi_k^{c}=c_k^{a}/a_k$, and $\phi_k^{d}=w_k^{a}/a_k$. The variability proxy is summarized separately as $\psi_k^{v}=w_k^{v}/(|\hat s_k(n^\star)|+\epsilon)$ and is used by the CRA to flag low-confidence locations where compensation should be conservative or escalated for human judgment.

\noindent\textbf{Compensation geometry under fixed tool tilt.}
The tool is mounted at a fixed tilt angle $\theta=25^\circ$ relative to the local reference axis used for compensation. Let $\Delta_k$ denote a signed equivalent correction along the tool axis required at pair $k$ for a selected component (for example, the drift component at $n^\star$, or a bounded correction derived from $c_k$). Under the fixed-tilt model, the corresponding axial and radial offset components are
$t_{l,k}=\Delta_k\cos\theta$ and $t_{r,k}=\Delta_k\sin\theta$.
The tools \texttt{rb\_compute\_tool\_length}, \texttt{rb\_compute\_radius\_offset}, and \texttt{rb\_compute\_pair\_tool\_comp} implement this mapping and return compensation vectors $\mathbf{t}_k=[t_{r,k},t_{l,k}]^{\top}$ that are then applied as radius and length offsets within the CAM environment. The CRA enforces bounds on $|t_{r,k}|$ and $|t_{l,k}|$ using knowledge graph-derived constraints and shop-floor limits before any recommendation is accepted.

\noindent\textbf{Knowledge retrieval for constraints and best practices.}
The knowledge graph tools \texttt{kg\_initial} and \texttt{kg\_retrieve} provide evidence-linked constraints, empirical rules, and known failure modes relevant to titanium machining and offset application. Retrieved constraints are represented as triple identifiers with supporting text spans. The CRA verifies that any constraint referenced in an accepted recommendation is backed by retrieved evidence identifiers, and otherwise triggers refinement or escalation.

\begin{table*}[t]
    \centering
    \caption{Main tool categories exposed to MAKA for rotor blade analysis and compensation. Implementation identifiers are provided to support reproducibility.}
    \label{tab:agent_tools}
    \begin{tabular}{p{3cm} p{6cm} p{6.0cm}}
        \toprule
        Category & Tool & Description \\
        \midrule
        Data loading
        & \texttt{compute\_inspection\_pairs}, \texttt{fetch\_inspection\_slices}
        & Load scan-based inspection measurements, construct paired pressure and suction statistics, and return cached normalized subsets for selected parts and pair keys. \\
        \\
        Statistics and indexing
        & \texttt{rb\_compute\_values}, \texttt{rb\_compute\_average}, \texttt{rb\_compute\_std\_dev}, \texttt{rb\_compute\_level}, \texttt{rb\_compute\_position\_in\_level}
        & Provide per-pair values and summary statistics and assign spanwise level and within-level position indices used for aggregation. \\
        \\
        Pathing projection
        & \texttt{rb\_compute\_pathing\_dev}
        & Project RMVM path-tracking error onto inspection keys in a common CAD frame and return per-surface pathing deviation $p_k$. \\
        \\
        Drift and variability proxies
        & \texttt{rb\_compute\_wear\_drift}, \texttt{rb\_compute\_process\_variability}, \texttt{rb\_compute\_residual\_systematic}
        & Estimate drift $b_k$ and derived drift magnitude $w_k^{d}$, residual dispersion $w_k^{v}$, and baseline systematic term $c_k$ from multi-part non-pathing deviations. \\
        \\
        Attribution metrics
        & \texttt{rb\_compute\_attribution\_fractions}
        & Compute numerically stable attribution fractions and variability indicators used for risk-aware verification and summarization. \\
        \\
        Compensation geometry
        & \texttt{rb\_compute\_tool\_length}, \texttt{rb\_compute\_radius\_offset}, \texttt{rb\_compute\_pair\_tool\_comp}
        & Map a selected signed correction $\Delta_k$ to axial and radial offsets under the fixed-tilt model and return $\mathbf{t}_k=[t_{r,k},t_{l,k}]^{\top}$. \\
        \\
        Knowledge retrieval
        & \texttt{kg\_initial}, \texttt{kg\_retrieve}
        & Build and query the machining knowledge graph and return evidence-linked constraints and best practices for verification and bounded action. \\
        \bottomrule
    \end{tabular}
\end{table*}

\section{Experimental Design}\label{sec:experimental_design}

\begin{figure*}
    \centering
    \includegraphics[width=\linewidth]{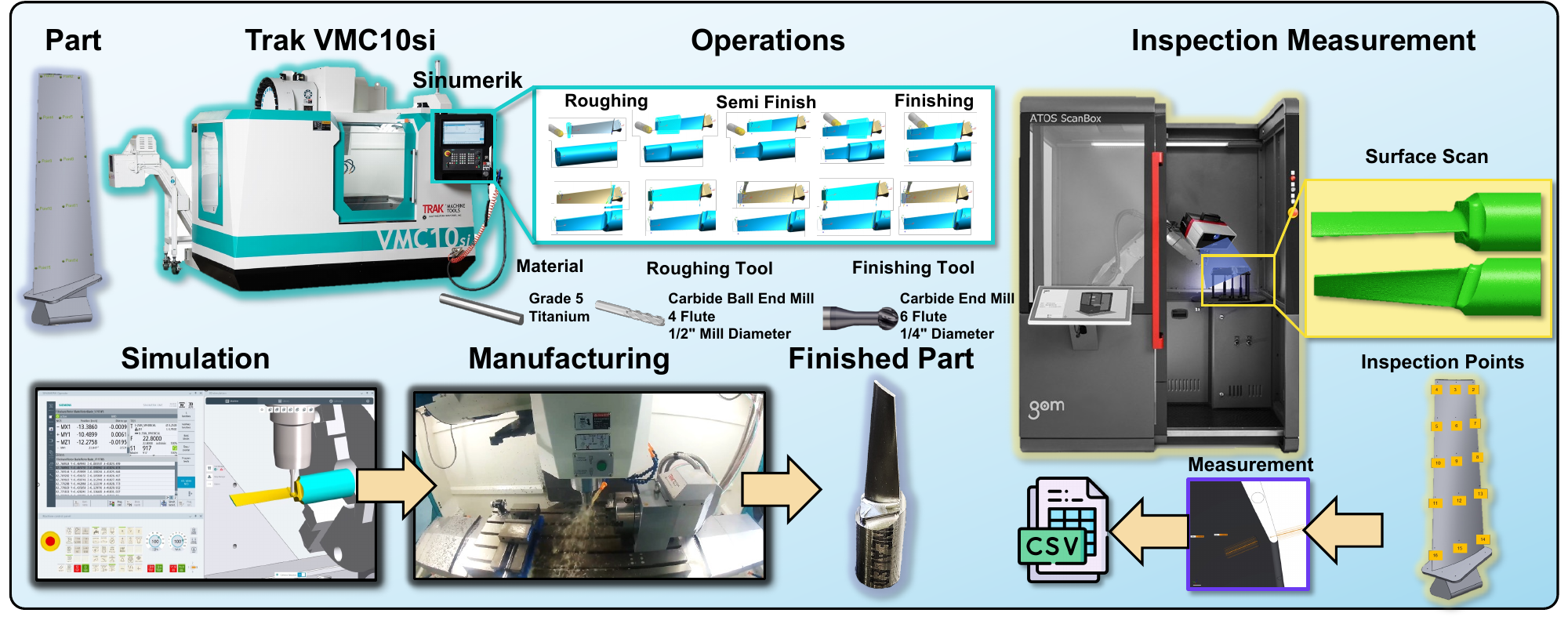}
    \caption{Illustration of the experimental workflow from part design, to manufacturing simulation, manufacturing using a 4-axis Trak VMC10si, and inspection measurements using a GOM ATOS ScanBox.}
    \label{fig:exp_design}
\end{figure*}

We evaluate the MAKA framework along four axes that map directly to the results section. First, we measure \emph{tool-orchestration reliability} on rotor blade analytics tasks, where the agent must select and sequence domain tools to compute deviation components and compensations from inspection data. Second, we isolate the marginal contribution of MAKA’s \emph{critic module} using paired runs with and without critic-mediated repair under controlled routing degradation. Third, we quantify the impact of \emph{knowledge graph (KG) grounding} on quantitative machining question answering using paired KG versus no-KG trials in both open-ended and multiple-choice formats. Finally, we present an \emph{end-to-end case study} in which inspection-derived boundary conditions and digital twin signals are converted into staged compensation actions that are verified in the digital twin before deployment.

\subsection{Manufacturing case study and inspection data}\label{subsec:exp_mfg_data}

The MAKA framework was evaluated using an experiment on the manufacturing of a rotor blade conducted at the Connecticut Center for Advanced Technology (CCAT). Figure~\ref{fig:exp_design} summarizes the overall fabrication, simulation, and inspection workflow. Ti-6Al-V titanium rods with a diameter of 1~1/4~in (31.75~mm) were used to manufacture a rotor blade with a length of 3~in (76.2~mm) designed at CCAT. Siemens NX CAM was used to generate baseline tool paths required to manufacture the rotor blade on a 4-axis Trak VMC10si.

\begin{figure*}
    \centering
    \includegraphics[width=\linewidth]{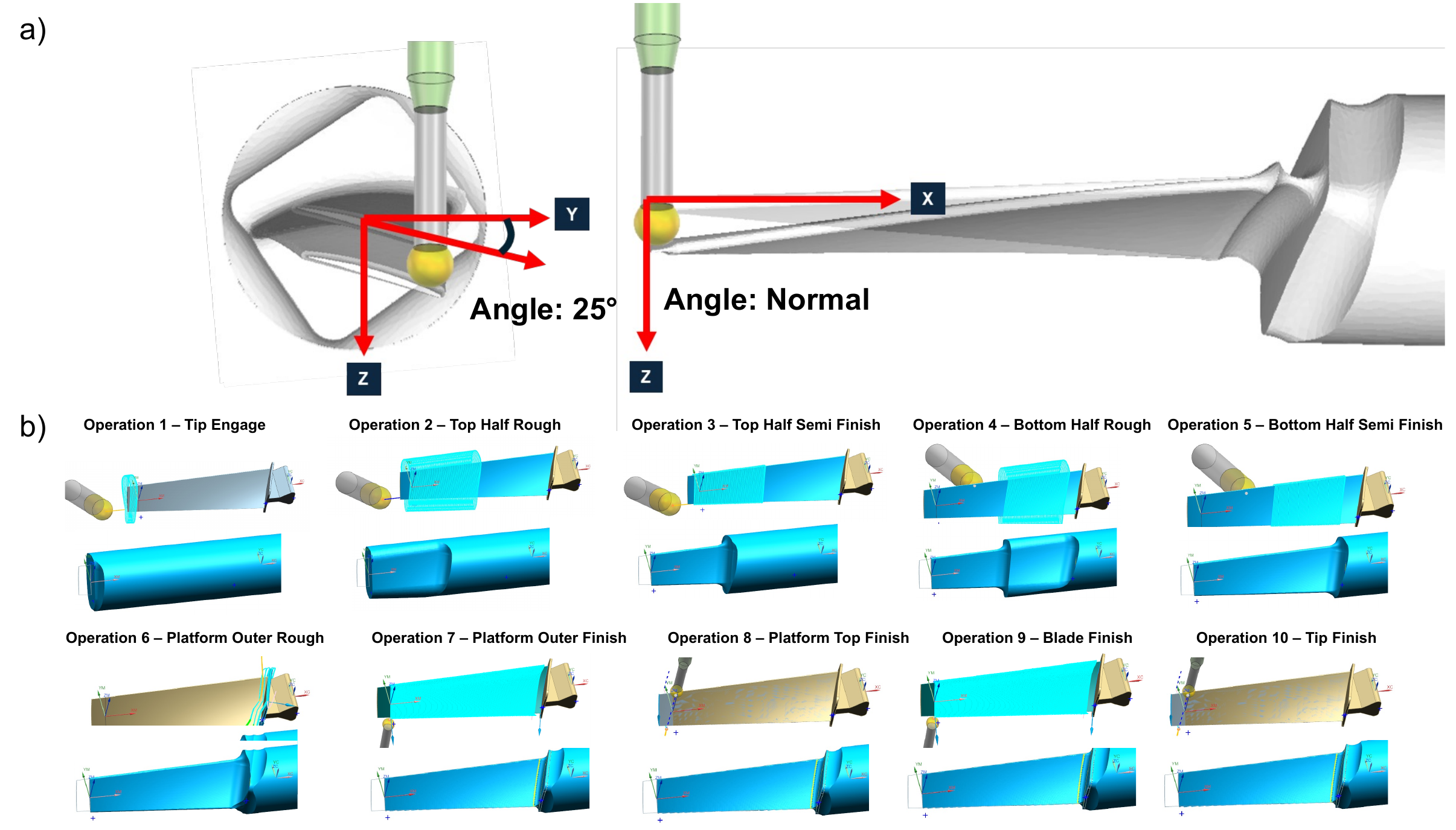}
    \caption{Experimental setup where (a) the spindle is set at an angle of $25^\circ$ relative to the blade and (b) all ten operations required to manufacture the rotor blade.}
    \label{fig:setup}
\end{figure*}

As shown in Figure~\ref{fig:setup}, ten operations are required to manufacture the blade: (1) tip engage, (2) top half rough, (3) top half semi-finish, (4) bottom half rough, (5) bottom half semi-finish, (6) platform outer rough, (7) platform outer finish, (8) platform top finish, (9) blade finish, and (10) tip finish. From these operations, four test cuts were selected to modify tool usage and update tool paths based on expert observations to minimize tool wear and improve surface finish. A four-flute 1/2~in TiAlN-coated carbide ball end mill was selected for all roughing and semi-finishing operations, and a six-flute 1/4~in solid carbide undercutting end mill was used for finishing operations. The tool was offset at an angle of 25 degrees relative to the part during machining (Figure~\ref{fig:setup}a).

A total of 16 rotor blades were created and measured for geometric deviation relative to the nominal CAD model using a GOM ATOS ScanBox. For each blade, surface scans were registered to the nominal model and deviation measurements were recorded at labeled inspection points. We defined 30 inspection points distributed across the blade surface and arranged them into matched pairs across the pressure and suction sides using a consistent \emph{pair-key} convention of the form $i+(i+15)$ (e.g., \texttt{2+17}, \texttt{3+18}, \ldots, \texttt{16+31}) so corresponding locations on the two sides could be directly compared. After measuring deviations, inspection data for all blades were exported to CSV and aggregated for subsequent analysis. These inspection deviations serve as the ground-truth geometric error signal that MAKA analyzes and attempts to reduce via compensation.

\subsection{Digital twin signals and deviation decomposition}\label{subsec:exp_dt_decomp}
MAKA forms an aggregate estimate of total geometric deviation by combining three complementary error signals with a residual term. The first signal captures cutting-induced workpiece deflection, the second captures machine path-tracking deviation, and the third captures systematic drift across parts as a proxy for wear-related evolution. Together, these signals provide a structured decomposition that separates deformation-driven effects, kinematic execution errors, and time-varying changes that accumulate over a production run.

Workpiece deflection is estimated using ThirdWave simulations that model how the blade deforms as the tool engages the pressure and suction sides under candidate process parameters such as spindle speed, feed rate, and surface feet per minute. The resulting deflection fields provide a physics-based estimate of deformation-induced deviation along the tool path, which allows MAKA to screen parameter sets toward reduced deflection sensitivity and to prioritize corrective actions in regions predicted to be most compliant.

Path-tracking deviation is estimated using Siemens Run MyVirtual Machine by comparing the predicted executed tool motion to the nominal G-code trajectory. These virtual machining results capture systematic tracking errors that arise from machine kinematics and controller behavior, and they provide a trajectory-following deviation field that MAKA can use to propose toolpath compensations aimed at counteracting repeatable execution bias.

Wear-related evolution is inferred directly from inspection measurements by tracking systematic drift in deviation patterns across the sequence of manufactured blades at fixed inspection pairs. Persistent, progressive changes across parts are treated as a proxy for wear-driven changes in effective tool geometry and engagement conditions. This drift-based signal is useful because it summarizes time-dependent behavior that is not captured by a single static simulation or a single part, and it provides a basis for wear-aware updates and conservative bounding when the process becomes less stable.

After accounting for predicted path-tracking effects and deflection trends and estimating drift-based evolution, any remaining systematic structure is treated as a residual compliance contribution. This residual term is intended to absorb unmodeled or partially calibrated effects and to prevent the decomposition from forcing all unexplained structure into one of the explicit contributors, which would otherwise lead to overconfident or misdirected compensations.

Within MAKA, these components are computed through the rotor blade analysis tool suite, including \texttt{rb\_compute\_pathing\_dev}, \texttt{rb\_compute\_surface\_dev}, \texttt{rb\_compute\_wear\_drift}, \texttt{rb\_compute\_process\_variability}, and \texttt{rb\_compute\_residual\_systematic}, and then combined into the decomposed representation used to generate compensation actions through modified tool paths and adjusted process parameters. Although direct point-to-point numerical agreement between simulation outputs and physical scans is not the primary objective, we observed strong topological alignment between the simulated and measured error distributions. ThirdWave consistently highlights the blade tip and leading edge as regions of maximum deflection, which matches the high-deviation zones observed in inspection across the 16 blades. This qualitative spatial agreement supports use of the digital twin as a directional guide for decision-making, while the residual term provides flexibility to reconcile absolute magnitudes and remaining systematic effects.

\subsection{Knowledge graph construction}\label{subsec:exp_kg}

The knowledge graph used in this experiment was constructed from a curated set of technical documents on machining Ti-6Al-4V (Table~\ref{tab:documents}), spanning cutting temperature and surface roughness optimization, tool wear mechanisms, sustainable cutting fluids, and machining deformation control in near-net-shaped jet engine blades. Together, these sources cover process parameters (cutting speed, feed, depth of cut, rake angle), performance measures (surface roughness, tool wear, tool life, cutting temperature), coolant strategies (including sustainable cutting fluids), and fixturing and deformation behavior in adaptive machining. The integrated KG contains 2701 triples over 3401 unique entities and 1151 unique relation types.

The KG was constructed automatically using the open-sourced \texttt{gpt-oss-120b} model served locally via an OpenAI-compatible HTTP API (LM Studio). Each paper was converted to Markdown and processed by a Python pipeline that splits text into overlapping windows (up to 1000 characters with a 500-character overlap). For each window, a prompt marks a central extraction region (\texttt{CHUNK\_MAIN}) and surrounding context windows, and specifies a strict five-field triple format:
\texttt{ENTITY\_1 <TAB> RELATIONSHIP\_TYPE <TAB> ENTITY\_2 <TAB> "RELATIONSHIP\_DESCRIPTION" <TAB> FIGURE\_REFERENCE}.
Triples are extracted from \texttt{CHUNK\_MAIN} only; surrounding context is used solely to resolve references to tables/figures. A post-processing stage enforces TSV consistency, repairs minor formatting errors, and propagates table references to related cell triples using schema triples. Triples are written per-document and then aggregated into the global KG used in subsequent experiments.

\begin{table*}[]
    \centering
    \caption{Overview of source documents and their corresponding knowledge graph sizes.}
    \scalebox{.95}{%
    \begin{tabular}{%
        p{4cm}
        p{8cm}
        >{\raggedright\arraybackslash}p{4cm}
    }
        \toprule
        \textbf{Document}
        & \textbf{Description}
        & \textbf{Knowledge Graph Size} \\
        \midrule
        Optimization of cutting temperature and surface roughness in CNC turning of Ti-6Al-4V alloy using response surface methodology~\cite{hossain2025optimization}
        & Introduced Response Surface Methodology to optimize cutting speed, feed rate, and depth of cut in dry CNC turning of Ti-6Al-4V alloy, aiming to minimize tool-chip interface temperature and surface roughness.
        & \makecell[tl]{
            \# of triples: 827  \\
            \# of unique entities: 978  \\
            \# of unique relations: 324
          } \\
        \midrule
        Evaluation of tool wears mechanism considering machining parameters and performance parameters for titanium alloy in turning operation on CNC~\cite{ingle2024evaluation}
        & Experimentally studies CNC turning of Ti-6Al-4V to evaluate how machining parameters (cutting speed, feed, depth of cut, and rake angle) affect tool wear, tool life, and surface roughness.
        & \makecell[tl]{
            \# of triples: 673 \\
            \# of unique entities: 802 \\
            \# of unique relations: 249
          } \\
        \midrule
        Sustainable green cutting fluid for interpreting optimization of process variables while machining on various CNC manufacturing systems---an experimental approach for exploring~\cite{ramena2025sustainable}
        & Compares several eco-friendly cutting fluids and finds that aerosol-mist neem oil is the most sustainable and effective choice for machining Ti-6Al-4V on CNC lathe and turn-mill systems, improving surface finish and reducing temperatures, especially in 4-axis turn-milling.
        & \makecell[tl]{
            \# of triples: 541 \\
            \# of unique entities: 707 \\
            \# of unique relations: 239
          } \\
        \midrule
        Analysis of machining deformation for adaptive CNC machining technology of near-net-shaped jet engine blade~\cite{wu2019analysis}
        & Develops and validates a rigid--flexible coupling fixture using PEEK-GF30 multi-point supports and an empirical cutting force model to analyze and control machining deformation of near-net-shaped TC4 jet engine blades in adaptive CNC machining.
        & \makecell[tl]{
            \# of triples: 660 \\
            \# of unique entities: 914 \\
            \# of unique relations: 339
          } \\
        \bottomrule
    \end{tabular}%
    }
    \label{tab:documents}
\end{table*}

\subsection{Evaluation protocols}\label{subsec:exp_protocols}

\subsubsection{Tool-use depth protocol (L1/L2/L3)}\label{subsec:exp_tooluse_eval}

To quantify tool-orchestration reliability on rotor blade analytics, we constructed 75 natural language questions that require computing deviation components and/or compensation values from inspection-derived signals. We group these into three tool-use depths: \textbf{L1} requires a single correct tool invocation; \textbf{L2} requires exactly two dependent tool calls with correct intermediate-result propagation; and \textbf{L3} requires three or more calls, stressing longer-horizon planning, state tracking, and dependency management. For each question, we compare the agent’s called-tool sequence (tools, order, and arguments) against a reference implementation derived from expert-designed CCAT scripts. We report tool-selection pass rate (\%) as the fraction of queries where the required tools are invoked with valid arguments and dependency ordering.

\subsubsection{Critic agent evaluation}\label{subsec:exp_critic_eval}

To quantify the contribution of the critic module within MAKA, we conducted a paired controlled evaluation in which each query was executed twice under matched conditions using the same base model family, question prompt, and initialization: one run with critic-mediated control and one run without critic mediation. In the critic-enabled condition, the execution path permits iterative validation and repair through critic decisions (e.g., retry or routing actions) before termination, whereas in the no-critic condition execution proceeds directly from routing and task agents to finalization without an explicit repair stage. This design isolates the marginal effect of critic intervention from confounds due to model choice or prompt variation. To test robustness to orchestration faults, we introduce controlled degradation at the first routing stage by deterministically removing a fixed fraction of required tool hints (default drop probability 0.3). Each trial is assessed against the query-specific required-tool specification (excluding infrastructure-only helper calls), and performance is recorded using tool-selection precision, recall, and F1. In addition to mean F1, we report improved-tool rate (fraction of paired queries where critic-enabled F1 exceeds no-critic F1), reduced-missing rate (fraction where critic reduces the count of missing required tools), and full-recovery rate over degraded cases (fraction of degraded trials where no-critic misses at least one required tool but the critic-enabled run recovers to zero missing required tools).

\subsubsection{Knowledge graph agent evaluation protocol}\label{subsec:exp_kg_eval}

Knowledge graph (KG) agent performance was evaluated in two linked phases using paired KG versus no-KG runs with the same base model family and identical question prompts: an open-ended quantitative phase and a multiple-choice phase derived from the same underlying items. The open-ended bank of 75 questions was programmatically constructed from the titanium machining sources and contains quantitative decision-style questions that require selecting operating points under thermal, roughness, wear, and throughput constraints, together with boundary and extrapolation checks, fixture-force reasoning, and speed-window compliance. Each item includes explicit ground-truth numeric targets, required terms, and scoring tolerances. The multiple-choice bank was generated from the same items by converting each ground-truth answer into one correct option and creating deterministic distractors via seeded numeric perturbations that remain outside acceptance tolerances. Performance reporting focuses on average numerical score versus average time per question for open-ended responses, and accuracy versus average time per question for multiple-choice responses, where time per question quantifies the inference-time overhead of KG retrieval relative to its corresponding gain in answer quality.

\section{Experimental Results}\label{sec:results}

MAKA was implemented in Python with a PostgreSQL backend used to store and query the machining knowledge graph. We report results in the same order as the experimental design: (i) tool-use reliability on rotor blade analytics tools, (ii) critic ablation under degraded routing, (iii) KG grounding ablation on quantitative machining QA, and (iv) an end-to-end case study demonstrating how inspection- and twin-derived signals are translated into staged compensation actions verified in the digital twin.

We first evaluate whether the agent can correctly select and sequence rotor blade analysis tools needed to compute deviation components and compensation values from inspection data. Queries require composing tools such as \texttt{rb\_compute\_surface\_dev}, \texttt{rb\_compute\_pathing\_dev}, \texttt{rb\_compute\_wear\_drift}, \texttt{rb\_compute\_process\_variability}, and \texttt{rb\_compute\_residual\_systematic} under increasing tool-chain depth (L1--L3). Tool-use pass rate (\%) is computed against expert reference scripts, counting a query as correct only when required tools are called with valid arguments and in dependency-consistent order.

\begin{figure*}
    \centering
    \includegraphics[width=\linewidth]{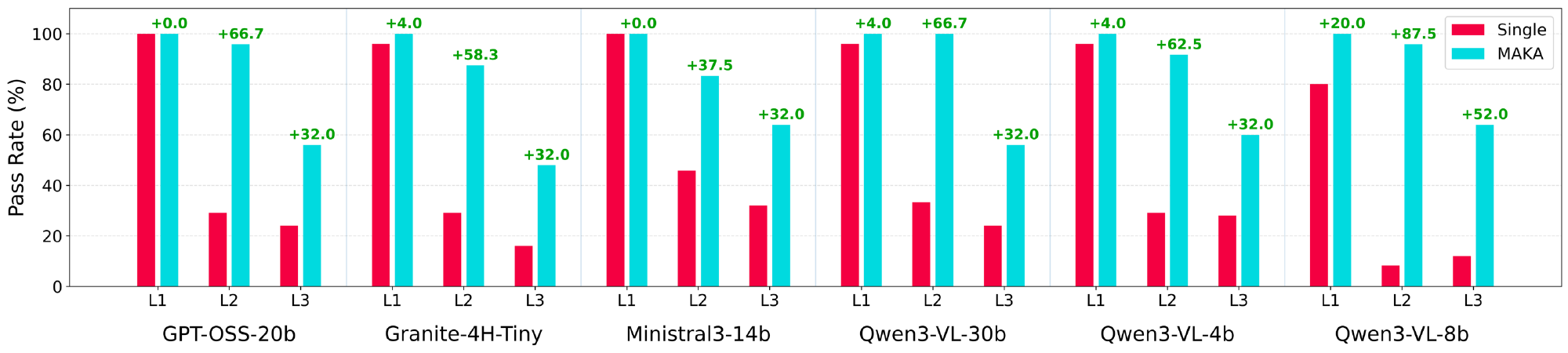}
    \caption{Comparison between MAKA and a single large language model in terms of pass rate (\%) for correct tool usage.}
    \label{fig:single_v_multi}
\end{figure*}

Figure~\ref{fig:single_v_multi} summarizes pass rates across three levels of tool-use depth (25 questions per level) and multiple open-source tool-trained models (GPT-OSS-20b, Granite-4H-Tiny, Ministral3-14b, Qwen3-VL-30b, Qwen3-VL-4b, and Qwen3-VL-8b). For L1 questions requiring a single tool invocation, both MAKA and single-LLM baselines achieve high pass rates, including perfect performance for stronger models such as GPT-OSS-20b and Ministral3-14b, while MAKA provides additional uplifts for weaker models (up to 20\% for Qwen3-VL-8b). As tool-use depth increases, differences become more pronounced: MAKA provides substantially larger gains on L2 (two dependent calls with correct intermediate propagation) and L3 (three or more calls with longer-horizon state tracking), where single-LLM baselines more frequently fail due to incorrect sequencing, invalid arguments, or loss of intermediate state. Overall, these results indicate that MAKA’s coordination and cross-step verification primarily reduce orchestration failures in compositional tool-use tasks rather than simply improving single-step tool execution.

\begin{table*}[t]
  \caption{Example of querying tool compensation using the MAKA framework.}
  \label{tab:tool-compensation}
  \renewcommand{\arraystretch}{1.1}

  {%
  \scriptsize
  \setlength{\tabcolsep}{3pt}

  \begin{tabularx}{\textwidth}{L}
    \toprule
    \textbf{Execution (CLI):}\\
    \rowcolor{gray!15}
    \begin{minipage}{.98\textwidth}
\ttfamily
Central Agent --- What would you like to do? \\
> load './Inspection\_Aggregated.csv' and give me compensation for parts 4 to 16 \\
    \end{minipage}\\[0.25em]
    \midrule

    \textbf{Model analysis:}\\
    The file \texttt{Inspection\_Aggregated.csv} is loaded, pairwise statistics are computed, and tool compensation values are returned for inspection pairs 4--16. Here, \texttt{Trc} denotes tool radius compensation and \texttt{Tlc} denotes tool length compensation:\\[0.25em]

    \rowcolor{gray!15}
    \begin{minipage}{.98\textwidth}
      \centering
      \begin{tabular}{@{}lll@{}}
        \toprule
        \textbf{Pair Key} & \textbf{Trc} & \textbf{Tlc} \\
        \midrule
        2+17  & 0.001164 & 0.002497 \\
        3+18  & 0.001030 & 0.002209 \\
        4+19  & 0.001063 & 0.002280 \\
        5+20  & 0.001104 & 0.002367 \\
        6+21  & 0.001138 & 0.002440 \\
        7+22  & 0.001253 & 0.002686 \\
        8+23  & 0.001466 & 0.003144 \\
        9+24  & 0.001290 & 0.002766 \\
        10+25 & 0.001356 & 0.002909 \\
        11+26 & 0.001513 & 0.003245 \\
        12+27 & 0.001475 & 0.003163 \\
        13+28 & 0.001556 & 0.003337 \\
        14+29 & 0.001670 & 0.003581 \\
        15+30 & 0.001528 & 0.003278 \\
        16+31 & 0.001620 & 0.003474 \\
        \bottomrule
      \end{tabular}
    \end{minipage}\\
    \bottomrule
  \end{tabularx}
  }
\end{table*}

Table~\ref{tab:tool-compensation} shows an example interaction in which the user requests compensation values over a selected part window. MAKA loads the aggregated inspection file, computes required pairwise statistics, and returns compensation values at each inspection pair key. This example is representative of the tool-use setting evaluated in Figure~\ref{fig:single_v_multi}: the agent must map a high-level request onto the correct rotor blade tools, respect dependencies among intermediate computations, and return numerically valid outputs in the expected format.

\begin{figure*}
    \centering
    \includegraphics[width=\linewidth]{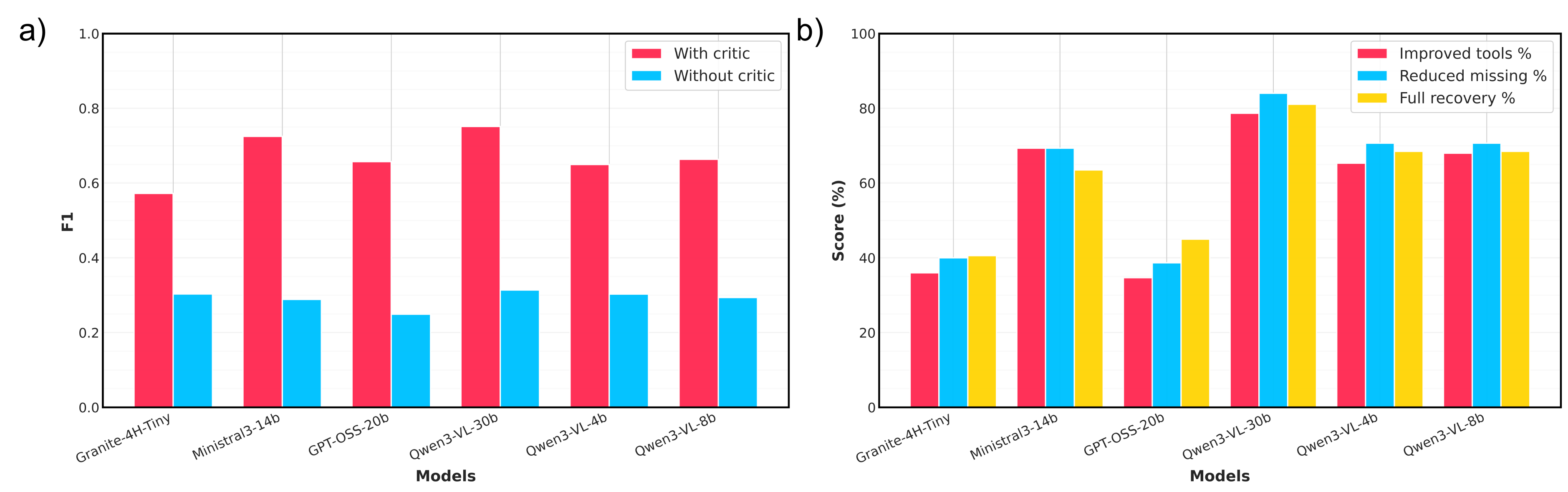}
    \caption{Critic agent ablation across six base models comparing critic-enabled and no-critic execution. (a) Average tool-selection F1 for each model with critic versus without critic. (b) Critic-value metrics from paired runs: improved tools (\%), reduced missing required tools (\%), and full recovery (\%) under degraded routing conditions.}
    \label{fig:critic_agent}
\end{figure*}

Figure~\ref{fig:critic_agent} reports the critic-agent evaluation described in Section~\ref{subsec:exp_critic_eval}. Each query is executed twice under matched conditions (critic-enabled vs no-critic), and deterministic first-step degradation removes a fraction of tool hints to probe recovery under orchestration stress. Across GPT-OSS-20b, Granite-4H-Tiny, Ministral3-14b, Qwen3-VL-30b, Qwen3-VL-4b, and Qwen3-VL-8b, each model is tested on 75 paired queries (450 paired trials total). Of these, 423 trials are degraded (some queries have no removable hints under the perturbation rule), enabling direct measurement of repair behavior under missing-dependency conditions. As shown in Figure~\ref{fig:critic_agent}a), the no-critic condition remains confined to a low F1 band between 0.2492 and 0.3137, whereas the critic-enabled condition rises to 0.5722 through 0.7511, and mean F1 increases from 0.2919 to 0.6697 across models. Absolute F1 gains are +0.2690 (Granite-4H-Tiny), +0.4360 (Ministral3-14b), +0.4079 (GPT-OSS-20b), +0.4374 (Qwen3-VL-30b), +0.3469 (Qwen3-VL-4b), and +0.3696 (Qwen3-VL-8b). Figure~\ref{fig:critic_agent}b) clarifies the mechanism: critic-enabled runs improve tool quality on 58.67\% of paired trials on average, reduce missing required tools on 62.22\%, and achieve full recovery on 61.19\% of degraded trials. Qwen3-VL-30b exhibits the strongest recovery profile (78.67\% improved tools, 84.00\% reduced missing, 81.08\% full recovery), while even the weakest recovery profile remains positive (GPT-OSS-20b at 34.67\%, 38.67\%, and 45.00\%). These results show that the critic functions as an explicit reliability layer that repairs missing dependencies, corrects incomplete tool chains, and prevents early routing errors from propagating into final answers.

\begin{figure*}
    \centering
    \includegraphics[width=\linewidth]{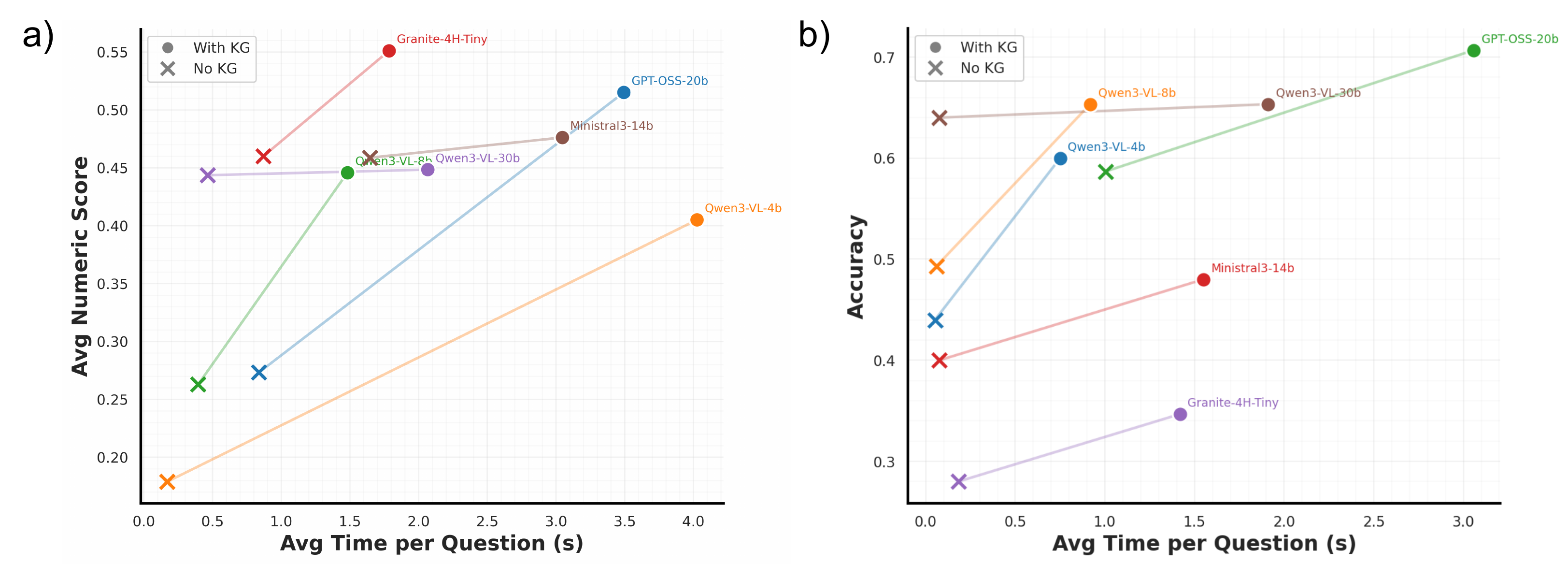}
    \caption{Knowledge graph evaluation across six base models using paired runs with and without KG retrieval. (a) Open-ended quantitative performance: average numerical score versus average time per question. (b) Multiple-choice performance: accuracy versus average time per question. Connected pairs show the shift from no-KG to with-KG, illustrating quality gains and inference-time overhead.}
    \label{fig:kg_agent}
\end{figure*}

Figure~\ref{fig:kg_agent} summarizes paired KG versus no-KG evaluation across the same six base models using 75 questions per model in each task format. In Figure~\ref{fig:kg_agent}a), every model moves from the no-KG point to a higher numeric score with KG enabled, increasing the cross-model mean from 0.3462 to 0.4736 (+0.1274), while average time per question increases from 0.7305~s to 2.6496~s (+1.9191~s). The largest open-ended gains occur for GPT-OSS-20b (+0.2418), Qwen3-VL-4b (+0.2262), and Qwen3-VL-8b (+0.1829). Granite-4H-Tiny improves moderately (+0.0911), while Qwen3-VL-30b and Ministral3-14b show smaller but still positive gains (+0.0048 and +0.0175), consistent with stronger no-KG baselines and diminishing marginal returns when parametric knowledge is already strong. In Figure~\ref{fig:kg_agent}b), the same trend appears for multiple-choice accuracy: mean accuracy rises from 0.4733 to 0.5733 (+0.1000), with average latency increasing from 0.2413~s to 1.6002~s (+1.3588~s). The largest MCQ uplifts are Qwen3-VL-4b and Qwen3-VL-8b (+0.1600 each), and GPT-OSS-20b both improves (+0.1200) and achieves the highest with-KG accuracy (0.7067). Overall, KG grounding provides reliable average gains in quantitative correctness and answer selection, with the largest benefits concentrated in models that most need external numeric evidence integration, and with a stable and interpretable latency tradeoff.

We close with an end-to-end case study that follows a complete MAKA loop from digital twin evidence and inspection measurements to a staged compensation plan that is verified in a high-fidelity digital twin prior to any shop-floor deployment. The intent is not to claim that any single simulation output numerically reproduces the scan data point by point. Instead, this case study shows that MAKA can extract directional guidance from physics-based simulation, reconcile that guidance with measured deviation structure and its evolution across parts, and translate the combined evidence into compensation actions that are auditable and bounded by uncertainty.

In this case study, MAKA receives three complementary inputs. ThirdWave provides deflection fields that indicate where the blade is most compliant and how sensitive that compliance is to changes in spindle speed and feed. The inspection scans provide the ground-truth geometric error signal relative to the nominal model at paired locations across the pressure and suction sides, revealing where systematic shape bias concentrates on the blade. Multi-part statistics derived from the inspection history summarize systematic evolution and dispersion at each inspection pair. MAKA uses these as proxies for wear-driven changes and process instability. These signals serve distinct roles. Simulation supplies a physics-based sensitivity map. Inspection supplies the realized error topology. Drift and variability provide a confidence and risk envelope that governs how aggressively compensation should be applied.

\begin{figure*}
    \centering
    \includegraphics[width=\linewidth]{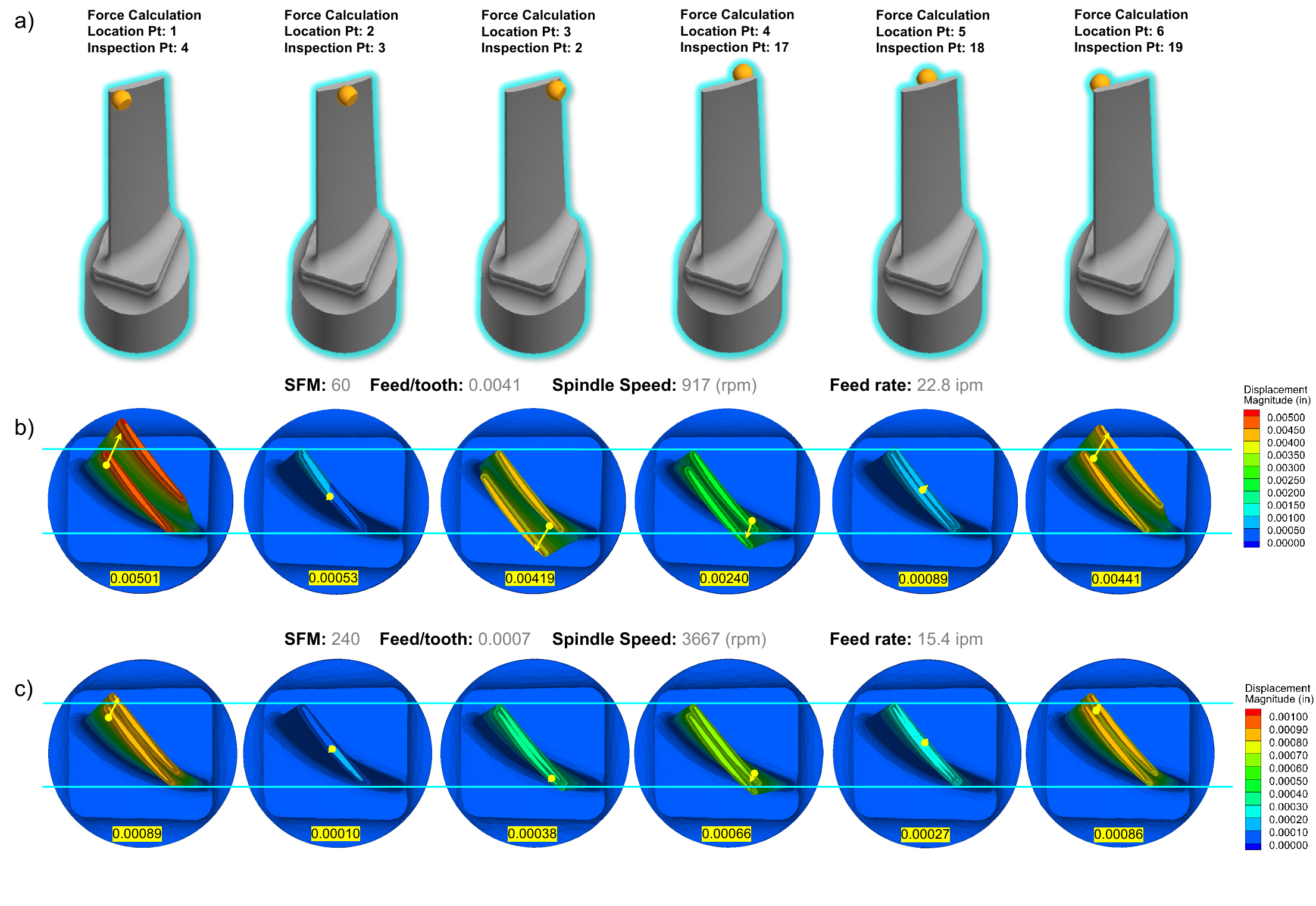}
    \caption{Deflection results. (a) The location where force calculations are applied. (b) Blade deflection for a spindle speed of 917~rpm and feed rate of 22.8~ipm. (c) Blade deflection for a spindle speed of 3667~rpm and feed rate of 15.4~ipm.}
    \label{fig:simulation}
\end{figure*}

Figure~\ref{fig:simulation} shows ThirdWave deflection simulations for two representative parameter sets selected to contrast an original Siemens NX recommendation with an alternative derived from expert machinist practice. Tool geometry, engagement conditions, and fixturing are held constant so the differences in the predicted fields are primarily attributable to spindle speed and feed. Beyond identifying a single minimum-displacement value, the simulations act as a screening mechanism that localizes deformation hot spots along the blade and establishes which parameter regimes are likely to reduce compliance-driven error. In thin-walled blade regions that behave like a cantilever, reductions in feed per tooth typically reduce cutting force peaks and can suppress elastic deflection, while increases in cutting speed can introduce thermal and wear considerations that must be evaluated in parallel. For this reason, MAKA treats the best-performing parameter set in the simulation as a candidate operating point that must be contextualized against stability, tool life, coolant strategy, and machine capability rather than as a universally optimal choice.

Table~\ref{tab:simulation-analysis} illustrates how MAKA converts the deflection visualization into an actionable recommendation in response to a natural language request. The agent identifies the minimum-displacement candidate across the shown parameter grid and then enumerates deployment considerations that are consistent with machining constraints. This step is important for closing the loop between physics and process engineering. The twin narrows the candidate space toward lower deformation sensitivity, and the KG-grounded considerations make explicit the failure modes that can negate those gains in practice, including accelerated wear at higher SFM, chatter margins, surface integrity effects, and coolant demands. MAKA therefore does not stop at ranking parameter sets. It surfaces the trade-offs that an engineer would normally check before committing to an update.

\begin{table*}[t]
  \caption{Example of analyzing simulation results using the MAKA framework.}
  \label{tab:simulation-analysis}
  \renewcommand{\arraystretch}{1.15}

  {%
  \scriptsize
  \setlength{\tabcolsep}{3pt}

  \begin{tabularx}{\textwidth}{L}
    \toprule
    \textbf{Execution (CLI)}\\
    \rowcolor{gray!15}
    \begin{minipage}{.98\textwidth}
\ttfamily
Central Agent - What would you like to do \\
> analyze the simulation results and give the process parameters with the lowest deviation and then list any potential considerations when using these parameters './data/Simulation\_results.png'\\
    \end{minipage}\\[0.25em]

    \rowcolor{gray!15}
    \begin{minipage}{\textwidth}
      \centering
      \includegraphics[width=0.75\linewidth]{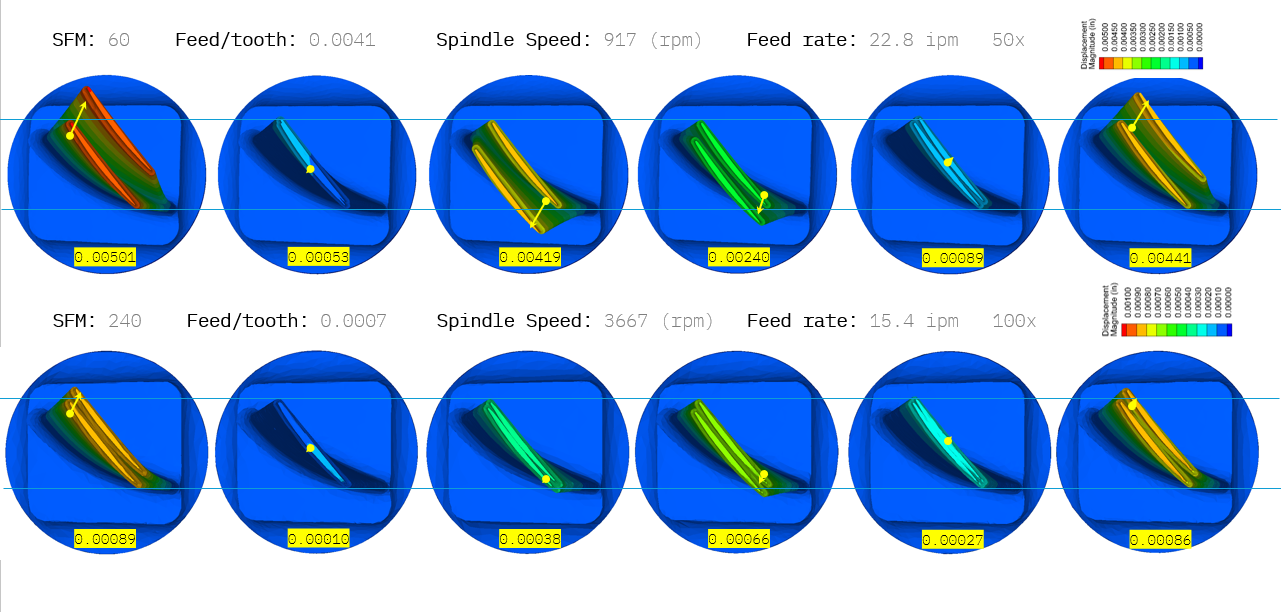}
    \end{minipage}\\
    \midrule

    \textbf{Model analysis}\\
    \rowcolor{gray!15}
    \begin{minipage}{.98\textwidth}
Final answer

Based on the simulation results shown in the image, the process parameters with the lowest deviation, meaning the smallest displacement magnitude, are as follows.

\textbf{Top Row (SFM 60)}
\begin{itemize}
  \item \textbf{Feed per tooth} 0.0041
  \item \textbf{Spindle Speed} 917 rpm
  \item \textbf{Feed Rate} 22.8 ipm
  \item \textbf{Lowest Displacement} 0.00089 (seen in the 5th image from the left)
\end{itemize}

\textbf{Bottom Row (SFM 240)}
\begin{itemize}
  \item \textbf{Feed per tooth} 0.0007
  \item \textbf{Spindle Speed} 3667 rpm
  \item \textbf{Feed Rate} 15.4 ipm
  \item \textbf{Lowest Displacement} 0.00010 (seen in the 2nd image from the left)
\end{itemize}

\textbf{Conclusion}
The parameters in the \textbf{bottom row (SFM 240)}, specifically \textbf{Feed per tooth 0.0007, Spindle Speed 3667 rpm, Feed Rate 15.4 ipm}, result in the lowest overall displacement magnitude (0.00010), indicating the most stable simulation outcome.

\medskip
\textbf{Potential considerations when using these parameters}

\begin{enumerate}
  \item \textbf{Tool wear and life.} Operating at higher SFM and spindle speed increases heat and can accelerate wear.
  \item \textbf{Machine tool capability.} Ensure the machine can sustain the spindle speed and feed without vibration.
  \item \textbf{Material removal rate.} Higher SFM changes MRR. Manage chip load and cutting forces accordingly.
  \item \textbf{Surface finish.} Lower deflection does not guarantee better finish. Verify with roughness checks.
  \item \textbf{Coolant requirements.} Higher cutting speeds may require more aggressive cooling.
  \item \textbf{Process stability.} Higher-speed regimes may be more chatter-sensitive if the system is not tuned.
\end{enumerate}
    \end{minipage}\\
    \bottomrule
  \end{tabularx}
  }
\end{table*}

While deflection screening addresses deformation sensitivity, inspection history is required to separate stable, compensable bias from evolving behavior that should be handled conservatively. Table~\ref{tab:tool-wear-analysis} shows how MAKA analyzes a wear-percentage envelope derived from inspection deviations over a selected part window. The key analytical signal is the envelope shape across inspection pairs. A narrow separation between minimum and maximum curves indicates that evolution is relatively consistent and can be treated as bounded drift. Widening separation indicates growing instability or mixed contributors that are not captured by a single drift trajectory. Envelope widening is a warning flag. It reduces confidence in applying aggressive offsets at those locations and instead favors bounded compensation, increased inspection frequency, or earlier tool-change thresholds. This drift and variability analysis therefore provides the gating logic that prevents MAKA from over-correcting regions where the process is no longer stable.

\begin{table*}[t]
  \caption{Example of analyzing tool wear results using the MAKA framework.}
  \label{tab:tool-wear-analysis}
  \renewcommand{\arraystretch}{1.15}

  {%
  \scriptsize
  \setlength{\tabcolsep}{3pt}

  \begin{tabularx}{\textwidth}{L}
    \toprule
    \textbf{Execution (CLI)}\\
    \rowcolor{gray!15}
    \begin{minipage}{.98\textwidth}
\ttfamily
Central Agent - What would you like to do \\
> analyze the tool wear percentage ranges and explain what can be shown by the plot (parts 4 to 20) \\
    \end{minipage}\\[0.25em]

    \rowcolor{gray!15}
    \begin{minipage}{\textwidth}
      \centering
      \includegraphics[width=0.75\linewidth]{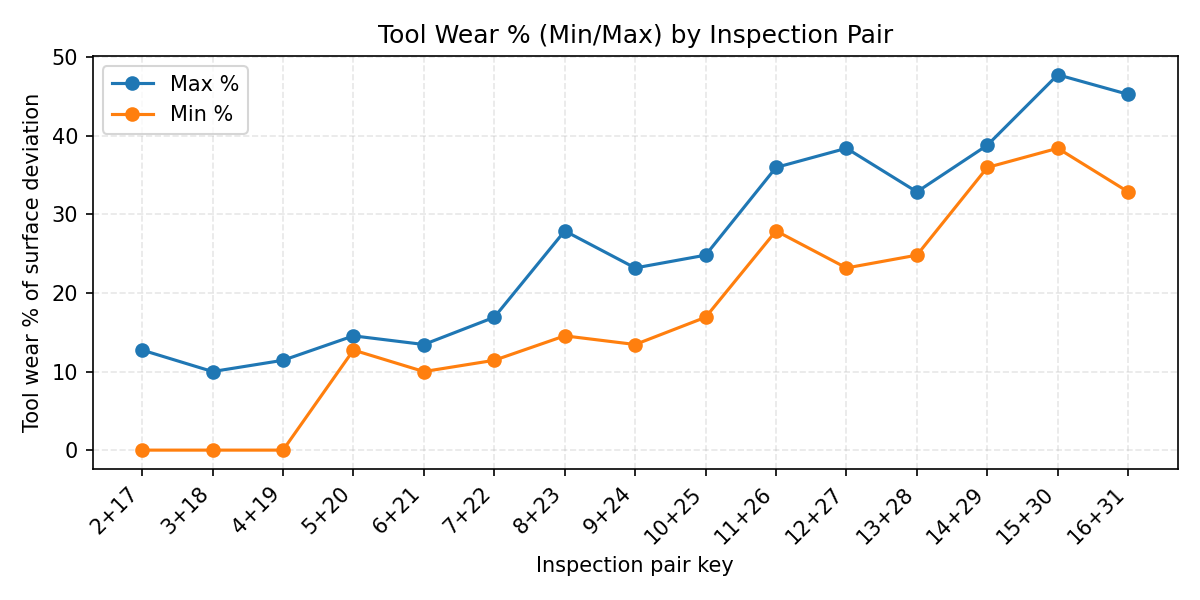}
    \end{minipage}\\
    \midrule

    \textbf{Model analysis}\\
    \rowcolor{gray!15}
    \begin{minipage}{.98\textwidth}
Wear percentage envelope metrics are computed and displayed for the selected part window. All values are calculated and non-NULL.

\medskip
\textbf{Final answer}

Based on the plot titled ``Tool Wear \% (Min/Max) by Inspection Pair'', here is what can be inferred.

\begin{enumerate}
  \item \textbf{What is being summarized}
        \begin{itemize}
          \item The vertical axis represents a normalized wear indicator expressed as a percentage contribution relative to the local deviation baseline used in the analysis.
          \item The horizontal axis lists inspection pair keys (for example ``2+17'' and ``3+18'') that correspond to fixed locations on the blade where multi-part deviations were tracked.
        \end{itemize}

  \item \textbf{How the two series should be interpreted under the drift and variability methodology}
        \begin{itemize}
          \item The two curves can be interpreted as an envelope over the evaluated part window that summarizes systematic evolution across parts.
          \item The separation between the curves provides an empirical indication of instability because a widening envelope implies that deviation evolution is not explained by a single tightly bounded drift trajectory.
        \end{itemize}

  \item \textbf{Key trends visible in the plot}
        \begin{itemize}
          \item \textbf{Drift envelope growth.} The envelope generally increases moving from left to right, indicating larger systematic evolution at later inspection locations.
          \item \textbf{Instability signature.} The gap between the envelope bounds becomes larger for later pairs, consistent with increasing process instability or unmodeled variation.
          \item \textbf{Peak location.} The largest envelope values occur near inspection pair ``15+30'', suggesting a wear-sensitive region for tighter monitoring or conservative offset bounds.
        \end{itemize}
\end{enumerate}

\medskip
\textbf{Summary.} The plot summarizes a drift-based wear envelope across inspection pairs and indicates where envelope widening suggests instability, supporting maintenance thresholds and conservative compensation decisions.
    \end{minipage}\\
    \bottomrule
  \end{tabularx}
  }
\end{table*}

In addition to producing numeric recommendations, MAKA is designed to provide explanations that a process engineer can audit. Table~\ref{tab:deflection-kg-analysis} illustrates this interpretability pathway. The agent takes a deflection trend summary and retrieves supporting causal relationships from the machining KG to produce a mechanism-level narrative for why deformation occurs. The retrieved factors include cutting deformation as the primary mechanism, sensitivity to cutting-force dynamics, low stiffness due to thin-walled cantilever geometry, stress concentration near clamping heads, and interactions between speed and wear. These factors justify why parameter tuning is a plausible corrective lever when deformation dominates, and they also highlight conditions under which the same parameter changes could increase risk, including higher wear rates or reduced stability margin. This makes the resulting compensation plan less of a black box because the system attaches physical and empirical rationale that explains why a given corrective action is being proposed.

\begin{table*}[t]
  \caption{Example of using a knowledge graph to explain rotor blade deflection.}
  \label{tab:deflection-kg-analysis}
  \renewcommand{\arraystretch}{1.15}

  {%
  \scriptsize
  \setlength{\tabcolsep}{3pt}

  \begin{tabularx}{\textwidth}{L}
    \toprule
    \textbf{Execution (CLI)}\\
    \rowcolor{gray!15}
    \begin{minipage}{.98\textwidth}
\ttfamily
Central Agent - What would you like to do \\
> causes of this deflection for the rotor blade \\
    \end{minipage}\\[0.25em]

    \rowcolor{gray!15}
    \begin{minipage}{\textwidth}
      \centering
      \includegraphics[width=0.75\linewidth]{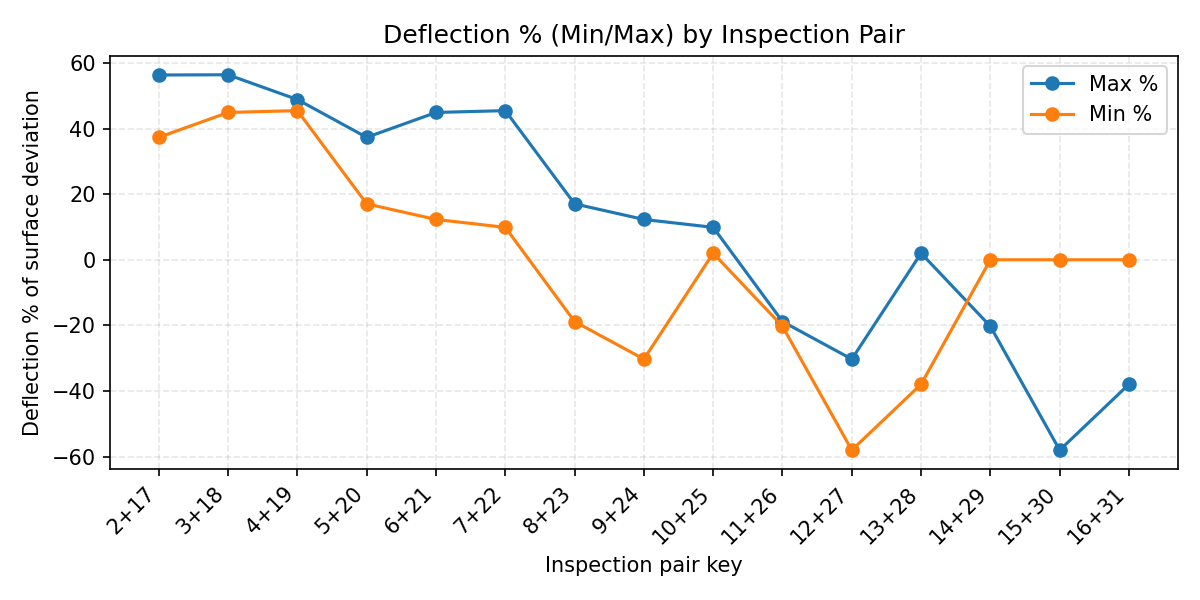}
    \end{minipage}\\
    \midrule

    \textbf{Model analysis (KG-enhanced)}\\
    \rowcolor{gray!15}
    \begin{minipage}{.98\textwidth}
Based on the retrieved knowledge and the analysis summary, the primary cause of rotor blade deflection is \textbf{cutting deformation}, which directly leads to the final deformation of the blade. This deformation is influenced by several factors.

\begin{enumerate}
  \item \textbf{Cutting force frequency.} Deflection increases significantly when the cutting force frequency exceeds 1500~Hz. At frequencies between 800 and 1500~Hz, the deformation is relatively small.
  \item \textbf{Blade structure.} The blade’s complex structure, including its thin-walled body, free surface, and large curvature changes, makes it susceptible to deformation. Its stiffness is low due to its cantilever beam structure.
  \item \textbf{Stress distribution.} Stress is higher near the clamping heads, which can lead to localized deformation or stress concentration.
  \item \textbf{Rotational speed and tool wear.} While increasing rotational speed slightly reduces cutting forces, it can also increase tool wear, which may indirectly affect blade deformation.
\end{enumerate}

The observed trends in the deflection chart, such as high deflection at the start (for example pair \texttt{2+17}) and more negative deflection later (for example pair \texttt{12+27}), likely reflect variations in cutting conditions such as force frequency, spindle speed, or tool wear across inspection pairs.
    \end{minipage}\\
    \bottomrule
  \end{tabularx}
  }
\end{table*}

\begin{figure*}
    \centering
    \includegraphics[width=.9\linewidth]{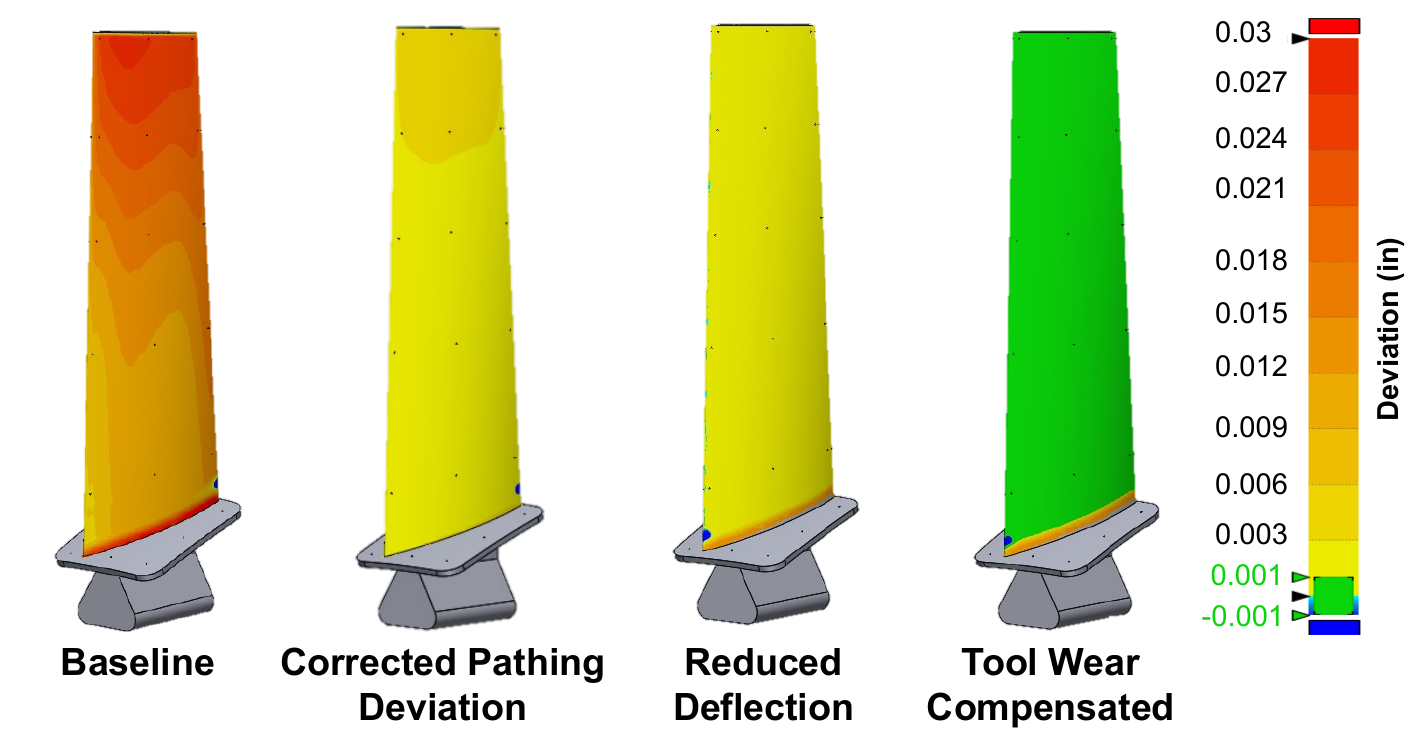}
    \caption{Successive MAKA-driven compensations for toolpath adjustment, deflection-focused parameter tuning, and tool-wear-aware updates, verified in the high-fidelity digital twin.}
    \label{fig:compensation}
\end{figure*}

Taken together, the deflection screening, drift and variability bounding, and KG-grounded explanations provide the evidence needed to justify staged compensation rather than a single monolithic correction. Toolpath updates are most effective when the observed error is spatially consistent and aligns with kinematic or path-tracking signatures. Parameter updates are most effective when the digital twin indicates high compliance sensitivity and the inspection error topology matches predicted deformation hot spots. Wear-aware bounding becomes critical when drift and variability indicate evolving behavior that would otherwise cause over-correction. This staged structure is reflected in the subsequent compensation verification (Figure~\ref{fig:compensation}), where successive steps reduce predicted deviation magnitude from the $10^{-2}$~in range toward the $10^{-3}$~in level by targeting distinct contributors in sequence, while using drift and variability as a safeguard against applying aggressive offsets in unstable regions.

\section{Conclusion}\label{sec:conclusion}

This paper introduces the multi agent knowledge analysis (MAKA) framework as a physics-grounded, LLM-driven decision-support layer for CNC manufacturing, in which a Central agent, Analysis agent, Knowledge Graph agent, and Critic agent translate natural language intent into verifiable tool use, evidence-linked retrieval, and risk-aware recommendations suitable for human approval prior to downstream CAM implementation. The architectural separation of intent routing, deterministic computation, knowledge retrieval, and verification is designed to support human--AI symbiosis in high-stakes manufacturing by reducing the cognitive burden of multimodal data synthesis while preserving auditability, bounded action, and provenance.

In the rotor blade case study, MAKA reliably orchestrated domain tools to align scan-based inspection deviations and virtual machining outputs at common inspection keys and to construct a physically interpretable decomposition consistent with the updated methodology. Specifically, pathing deviation was obtained from Run MyVirtual Machine projections, systematic evolution across parts was captured through a drift-based wear proxy estimated from multi-part inspection sequences, and the remaining systematic term was treated as a residual compliance contribution rather than being uniquely attributed to deflection. A variability proxy was also computed to characterize instability and to inform conservative recommendation and escalation when confidence was insufficient. This tool-grounded decomposition reproduces the structure of expert workflows while making the assumptions explicit and inspectable by the human operator and by the Critic agent verification gate.

Beyond the case study demonstration, controlled ablations show that MAKA's multi-agent decomposition is necessary for reliable performance once tasks require compositional tool use. In the multi-agent versus single-LLM evaluation, single-LLM baselines performed competitively on L1 queries that require only one tool call, but their performance degraded as soon as questions required dependent tool chains. MAKA delivered substantially larger gains on L2 and L3 queries, including uplifts up to 87.5\% on L2 and at least 32\% on L3, where failures are dominated by orchestration errors rather than missing domain knowledge. These results indicate that separating intent routing, deterministic computation, and stateful multi-step execution is not an implementation convenience. It directly reduces invalid argument bindings, incorrect call ordering, and intermediate-state loss, which are precisely the failure modes that make single-agent tool use brittle in manufacturing decision-support settings.

Critic and knowledge graph ablations further clarify why additional specialized agents are required to make this structure robust and trustworthy under realistic orchestration noise. Under controlled routing degradation, critic-enabled runs increased mean tool-selection F1 from 0.2919 to 0.6697 across models and achieved full recovery on 61.19\% of degraded trials, demonstrating that explicit verification and repair is critical for preventing early routing faults from propagating into unusable or unsafe outputs. Similarly, knowledge graph grounding produced consistent improvements in quantitative machining answers, raising mean open-ended score from 0.3462 to 0.4736 and mean multiple-choice accuracy from 0.4733 to 0.5733, while adding a predictable latency overhead that remains negligible relative to machining cycle times. Taken together, these ablations show that the additional agents are not optional performance enhancements but explicit reliability mechanisms. Multi-agent orchestration provides the structure needed for correct tool chaining, the critic provides a recovery layer under missing or degraded dependencies, and the knowledge graph agent supplies evidence-linked numeric grounding and constraint-aware caveats that reduce hallucination risk and improve decision quality.

By coupling these decomposed contributors with geometric compensation relationships under fixed tool tilt and with physics-based deflection simulation, MAKA generated actionable toolpath and parameter compensation candidates compatible with NX CAM. The effect of successive compensations was evaluated within a high-fidelity digital twin toolchain, providing a verified pre-deployment baseline for risk-aware human decision-making rather than a claim of physical closed-loop correction. Across the staged compensation sequence, the predicted deviation field progressed from large localized errors on the order of $10^{-2}$~in to a predominantly near-nominal state within approximately $\pm 0.001$~in over most of the surface, demonstrating that coordinated, tool-grounded reasoning can drive the simulated process toward a high-accuracy operating regime while preserving interpretability and traceability. The complete multi-agent reasoning and retrieval loop executes in 4.3~s. Relative to aerospace machining cycle times that are typically minutes to hours for complex rotor blades, this latency is negligible, rendering the computational overhead effectively inconsequential. As a result, MAKA can be invoked between machining stages to assimilate inspection data or process telemetry and inform subsequent decisions without inducing machine idle time, enabling practical inter-process decision support in production settings.


\section*{Declaration of Competing Interest}
The authors declare that they have no known competing financial interests or personal relationships that could have appeared to influence the work reported in this paper.

\section*{Acknowledgments}
This work was supported under Cooperative Agreement W56HZV-21-2-0001 with the U.S. Army DEVCOM Ground Vehicle Systems Center (GVSC), through the Virtual Prototyping of Autonomy Enabled Ground Systems (VIPR-GS) program; the National Science Foundation (grant no. 2434519); and the Department of Energy (grant no. DE-EE0011029). The authors gratefully acknowledge the valuable contributions from the Connecticut Center for Advanced Technology (CCAT) for this research.


\bibliographystyle{elsarticle-num-names} 
\bibliography{references}

\end{document}